\documentclass[10pt,journal]{IEEEtran}
\usepackage{nccmath}
\usepackage{cases} 
\usepackage{amssymb}
\usepackage{amsmath}
\usepackage{algorithmic}
\usepackage{algorithm}
\usepackage{cite}
\usepackage{url}
\usepackage{xcolor}
\usepackage{color} 
\usepackage{cite,graphicx,amsmath,amssymb}
\usepackage{subfigure}
\usepackage{citesort}
\usepackage{fancyhdr}
\usepackage{mdwmath}
\usepackage{mdwtab}
\usepackage{caption}
\usepackage{amsthm}
\usepackage{setspace}
\usepackage{xspace}
\usepackage{graphicx}
\usepackage{threeparttable}
\usepackage{makecell}
\usepackage{booktabs}
\usepackage{pifont}
\newcommand{\cmark}{\ding{52}}

\DeclareMathOperator*{\argmax}{arg\,max}
\usepackage[acronym,nogroupskip,nonumberlist,nopostdot]{glossaries}
\makeglossaries
\newacronym{aod}{AOD}{angle of departure}
\newacronym{aoa}{AOA}{angle of arrival}
\newacronym{snr}{SNR}{signal-to-noise ratio}
\newacronym{6g}{6G}{sixth-generation}
\newacronym{5g}{5G}{fifth-generation}
\newacronym{sre}{SRE}{smart radio environment}
\newacronym{stars}{STAR-RISs}{simultaneously transmitting and reflecting reconfigurable intelligent surfaces}
\newacronym{star}{STAR-RIS}{simultaneously transmitting and reflecting reconfigurable intelligent surface}
\newacronym{mimo}{MIMO}{multiple-input multiple-output}
\newacronym{cscg}{CSCG}{circularly symmetric complex Gaussian}
\newacronym{los}{LoS}{line-of-sight}
\newacronym{em}{EM}{electromagnetic}

\DeclareMathOperator{\diag}{diag}

\newcommand{\bm}[1]{\mbox{\boldmath{$#1$}}}

\newtheorem{theorem}{Theorem}

\newtheorem{lemma}{Lemma}

\newtheorem{property}{Property}

\newtheorem{proposition}{Proposition}

\allowdisplaybreaks
\makeatletter
\def\ScaleIfNeeded{%
\ifdim\Gin@nat@width>\linewidth \linewidth \else \Gin@nat@width
\fi } \makeatother

\begin{document}

\title{STAR-RIS in Cognitive Radio Networks}

\author{
Haochen~Li,~\IEEEmembership{Graduate Student Member,~IEEE,}
Yuanwei~Liu,~\IEEEmembership{Fellow,~IEEE,}
Xidong~Mu,~\IEEEmembership{Member,~IEEE,}
Yue~Chen,~\IEEEmembership{Senior Member,~IEEE,}
Pan~Zhiwen,~\IEEEmembership{Member,~IEEE,}
Xiaohu~You,~\IEEEmembership{Fellow,~IEEE}
\thanks{Part of this article has been accepted by the IEEE Global Communications Conference, in 2024~\cite{Li}. \textit{(Corresponding author: Pan Zhiwen; Yuanwei Liu.)}}
\thanks{Haochen~Li, Pan~Zhiwen, and Xaiohu~You are with National Mobile Communications Research Laboratory, Southeast University, Nanjing 210096, China, and also with Purple Mountain Laboratories, Nanjing 211100, China (email: lihaochen@seu.edu.cn, pzw@seu.edu.cn, xhyou@seu.edu.cn).}

\thanks{Yuanwei Liu is with the School of Electronic Engineering and Computer Science, Queen Mary University of London (QMUL), E1 4NS London, U.K., and also with the Department of Electronic Engineering, Kyung Hee University, Yongin-si, Gyeonggi-do 17104, South Korea (e-mail: yuanwei.liu@ qmul.ac.uk).}

\thanks{Xidong~Mu is with the Centre for Wireless Innovation (CWI), Queen's University Belfast, Belfast, BT3 9DT, U.K. (e-mail: x.mu@qub.ac.uk).}

\thanks{Yue~Chen is with the School of Electronic Engineering and Computer Science, Queen Mary University of London, London E1 4NS, U.K. (e-mail: yue.chen@qmul.ac.uk).}

}
\maketitle
\begin{abstract}
The development of sixth-generation (6G) communication technologies is confronted with the significant challenge of spectrum resource shortage. To alleviate this issue, we propose a novel simultaneously transmitting and reflecting reconfigurable intelligent surface (STAR-RIS) aided multiple-input multiple-output (MIMO) cognitive radio (CR) system. Specifically, the underlying secondary network in the proposed CR system reuses the same frequency resources occupied by the primary network with the help of the STAR-RIS. The secondary network sum rate maximization problem is first formulated for the STAR-RIS aided MIMO CR system. The adoption of STAR-RIS necessitates an intricate beamforming design for the considered system due to its large number of coupled coefficients. The block coordinate descent method is employed to address the formulated optimization problem. In each iteration, the beamformers at the secondary base station (SBS) are optimized by solving a quadratically constrained quadratic program (QCQP) problem. Concurrently, the STAR-RIS passive beamforming problem is resolved using tailored algorithms designed for the two phase-shift models: 1) For the \emph{independent phase-shift model}, a successive convex approximation-based algorithm is proposed. 2) For the \emph{coupled phase-shift model}, a penalty dual decomposition-based algorithm is conceived, in which the phase shifts and amplitudes of the STAR-RIS elements are optimized using closed-form solutions. Simulation results show that: 1) The proposed STAR-RIS aided CR communication framework can significantly enhance the sum rate of the secondary system. 2) The coupled phase-shift model results in limited performance degradation compared to the independent phase-shift model.
\end{abstract}


\section{Introduction}
With the exponential growth of data traffic in recent years, the forthcoming sixth generation (6G) communications are envisioned to provide Terabits per second (Tbps) data rates, meeting diverse service requirements despite the challenges posed by limited spectrum resources~\cite{8766143}. Cognitive radio (CR) stands out as an effective technology aimed at bolstering the spectrum efficiency (SE) of communication systems through spectrum sharing. In CR systems, both the primary network and the secondary network reuse the same spectrum resources~\cite{5783948}. {In this work, we investigate the underlay CR, where the primary users (PUs) within the primary network have a higher priority for spectrum utilization compared to the secondary users (SUs) in the secondary network.} {{Specifically, the methods of spectrum sharing in CR systems can be categorized into interweave, overlay, and underlay modes~\cite{wang2010advances}. In the interweave mode, also known as opportunistic spectrum access, SUs sense the spectrum to identify spectrum holes and use these gaps for transmission~\cite{9371416,9687460}. In the overlay mode, SUs are allowed to transmit simultaneously with PUs by using advanced signal processing and coding techniques~\cite{liang2017cooperative}.}} {To ensure the coexistence of the two networks, it is imperative to restrict interference from the secondary network on all PUs below a specified threshold.} Consequently, effective interference management in CR systems holds paramount significance, leading to various interference management approaches~\cite{7111366,5280202,5403537}.

{Note that conventional CR system designs focus on interference management under the given wireless propagation environment. Recently, reconfigurable intelligent surfaces (RISs) have emerged as a nascent technology. A RIS consists of a large number of passive elements. It can manipulate the propagation environment by intelligently reflecting signals in desired directions~\cite{9090356,10158690,10183797}. This ability to perform passive beamforming allows signals to be directed toward intended receivers while reducing leakage toward unintended directions~\cite{9690635}. Additionally, a RIS can create destructive signals to cancel unwanted signals~\cite{9681803}. These capabilities offer a novel approach to addressing interference issues in CR systems~\cite{9140329}. Therefore, RIS-aided CR communication is regarded as a promising solution.}
\subsection{Prior Works}
\subsubsection{RIS aided CR systems}
Considering that RISs can enhance the useful signal for SUs and mitigate interference to PUs in CR systems, substantial research has been devoted to exploring RIS-aided CR systems. In~\cite{9235486}, the authors proposed a RIS-aided {multiple-input single-output (MISO) CR system} with a single SU, presenting a joint beamforming strategy for maximizing SU SE. {However, the algorithm proposed in this work cannot be easily extended to CR systems with multi-antenna SUs or multiple SUs. To fill this research gap,} the authors of~\cite{9599553} investigated a RIS-aided {multiple-input multiple-output (MIMO) CR system} with multiple SUs, focusing on maximizing the sum rate of SUs using an alternating optimization (AO)-based algorithm. {Work~\cite{9599553} requires perfect channel state information (CSI) at the BS to carry out beamforming design. Recognizing the challenge of obtaining CSI for RIS-related channels,} a robust joint beamforming design algorithm based on imperfect CSI is proposed for RIS-aided MIMO CR systems in~\cite{zhang2021robust}. {In full-duplex communication, both sending and receiving can happen simultaneously, potentially doubling the data throughput for the same bandwidth. To further improve the system SE,} full-duplex CR systems were investigated to maximize the weighted sum rate of both uplink and downlink SUs in~\cite{9183907}.
\subsubsection{STAR-RIS aided CR systems}
The aforementioned works focus on conventional RIS aided CR systems. {However, the primary and secondary networks must be situated on the same side of the RIS in order to use traditional RIS to support the CR system. This geographical restriction may not always be satisfied in real communication systems, especially for mobile users in wireless communication systems. As a remedy, the concept of simultaneously transmitting and reflecting RISs (STAR-RISs) can be used to address this limitation~\cite{10550177}.} The STAR-RIS can partition the incoming signal into both the transmission region and the reflection region using various working protocols, i.e., the time switching (TS) protocol, the mode switching (MS) protocol, and the energy splitting (ES) protocol~\cite{mu2021simultaneously}. This approach enables full-space coverage and offers additional degrees of freedom (DoFs) for beamforming design~\cite{9437234}. To further enhance the performance of CR systems, some initial efforts~\cite{10347404,10345673,10263782,10093979} have explored the use of STAR-RIS to assist CR systems in communication security, non-terrestrial vehicle networks, and Internet of things (IoT) networks, respectively. The authors in~\cite{10347404} investigated the communication security for the STAR-RIS assisted nonorthogonal multiple access (NOMA) CR system. Based on the Gaussian-Chebyshev quadrature, the analytical expressions of the outage probability and intercept probability for IoT users were derived. The authors in~\cite{10345673} investigated the STAR-RIS aided CR system with the hybrid automatic repeat request scheme, where the users with security demand are paired with the users with low quality of service demand to carry out NOMA. With derived analytical expressions, the trade-off between paired users was presented. The authors in~\cite{10263782} investigated adopting STAR-RIS aided CR system for consumer IoT networks and analyzed the energy efficiency for IoT users. The authors in~\cite{10093979} explored a STAR-RIS aided non-terrestrial vehicle network empowered by NOMA and assessed the outage performance of all secondary vehicles. {Nonetheless, these works primarily focused on the performance analysis of STAR-RIS aided CR systems with the simple single-transmit and single-receive antenna configuration. Considering multi-antenna BS,} the authors in~\cite{wen2022star} investigated the beamforming design of the STAR-RIS aided MISO CR system, where the secondary network power consumption minimization problem is solved with the STAR-RIS working under the TS protocol. { Works~\cite{10347404,10093979,10345673,wen2022star} are limited to cases where both the reflection and transmission spaces of the STAR-RIS have single SU. This limitation often cannot be met in practical CR systems.}

\subsection{Motivations and Contributions}
The deployment of STAR-RISs not only brings full space coverage for the CR system but also necessitates a more intricate beamforming design. This complexity arises due to the introduction of a larger number of adjustable coefficients compared to conventional RISs. {Although the authors of~\cite{wen2022star} have investigated the beamforming design for the STAR-RIS aided CR system, it should be noted that the employed TS working protocol for STAR-RIS in this study does not allow simultaneous service provision for users on both sides of the STAR-RIS. To achieve actual simultaneous service for users across the full space, more advanced working protocols, such as ES protocol, should be considered.}

In contrast to the TS STAR-RISs, whose elements operate exclusively in either reflection or transmission mode, all STAR elements in the ES STAR-RISs operate in the reflection and transmission modes. In this circumstance, the reflection and transmission coefficients of the same STAR element should be jointly considered, as they represent concurrent parameters for the same hardware component. There are four adjustable coefficients for each STAR element, i.e., transmission amplitude, reflection amplitude, transmission phase, and reflection phase. The {independent phase-shift model} and the {coupled phase-shift model} have been proposed to characterize the relation between them~\cite{9774942}. The latter model introduces an additional constraint on the phase coefficients for the same STAR element, which accounts for hardware limitations in STAR elements. Tailored passive beamforming designs are necessary when utilizing various STAR-RIS models.

Against this background, we investigate the STAR-RIS aided MIMO underlay CR system in this work. The performance of CR systems assisted by STAR-RIS is comprehensively assessed using both the {independent phase-shift model} and the {coupled phase-shift model}. Contributions of this paper can be summarized as follows:
\begin{itemize}
\item A STAR-RIS aided MIMO CR communication framework is proposed where an SBS communicates with SUs over the same frequency band occupied by the primary network. The sum rate maximization problem for SUs is formulated to optimize beamformers at the SBS and coefficients of the STAR-RIS under the SBS power constraint, the interference constraints, and the STAR-RIS coefficients constraint.

\item For the {independent phase-shift model}, we first reformulate the optimization problem using the Lagrangian dual transform (LDT) and the complex quadratic transform (CQT). {Subsequently, we propose a block coordinate descent (BCD)-based algorithm to solve the optimization problem.} {With given STAR-RIS configuration, the beamforming design at the SBS is optimized by solving a quadratically constrained quadratic program (QCQP) problem. With given SBS beamformers, the STAR-RIS coefficients are optimized with a successive convex approximation (SCA)-based algorithm.}

\item For the {coupled phase-shift model}, the corresponding passive beamforming problem is solved using the penalty dual decomposition (PDD)-based technique in a two-loop iteration. The Lagrangian dual vector and penalty factor are updated in the outer iteration loop. The STAR-RIS's amplitudes and phase shifts are alternatively optimized with closed-form solutions in the inner iteration loop.

\item Our simulation results validate: 1) the superior performance of the STAR-RIS aided CR communication framework compared to other benchmark schemes; and 2) the degradation in performance due to coupled phase shifts in STAR-RIS aided CR systems is limited compared to the independent phase STAR-RIS.
\end{itemize}

\subsection{Organization and Notations}
The remainder of this manuscript is organized as follows: Section II outlines the system architecture, STAR-RIS models, and signal model of the STAR-RIS-aided CR system. Following this, Section III presents the problem formulation for optimizing the sum rate of the STAR-RIS-assisted CR system, employing a BCD-based approach to address the optimization challenges associated with the {independent phase-shift model}. In Section IV, a PDD-based approach is introduced to overcome the challenges posed by the {coupled phase-shift model}. Section V presents simulation results that validate the effectiveness of the proposed strategies. Finally, Section VI provides the conclusion for this paper.

Lowercase, uppercase, and lowercase bold letters stand for scalars, vectors, and matrices, respectively. The space of $M \times K$ dimensional complex matrices is represented by the notation $\mathbb{C}^{M \times K}$. $(\cdot)^\mathrm{T}$, $(\cdot)^\mathrm{*}$, and $(\cdot)^\mathrm{H}$ are the superscripts for the transpose, conjugate, and conjugate transpose operations, respectively. A vector consisting of the primary diagonal elements of the matrix $\mathbf{A}$ is denoted by the notation $\text{diag}(\mathbf{A})$. $\text{tr}(\mathbf{A})$ and $\text{rank}(\mathbf{A})$, respectively, represent the trace and rank of matrix $\mathbf{A}$. $\Re(\cdot)$ and $\angle(\cdot)$ are used to extract the real component and angle of a complex number, respectively. The complex number's absolute value, the Euclidean norm, and the infinity norm are denoted by the notations $|\cdot|$, $|\cdot|_2$, and $|\cdot|_\infty$, respectively. Calligraphic letters such as $\mathcal{A}$ are used to represent sets. {A list of important variables is given in Table~\ref{Symbols}.}
\begin{table}[!t]
\caption{{List of Important Variables}}\label{Symbols}
\centering
\begin{tabular}{|c||c|}
\hline
Symbol & Definition\\
\hline
$\mathbf{v}_i, i\in\{t,r\}$ & STAR-RIS coefficients vectors\\
\hline
$\mathbf{\Phi}_i, i\in\{t,r\}$ & STAR-RIS coefficients matrices\\
\hline
$\mathbf{w}_l, l\in\mathcal{L}$ & SBS beamformers\\
\hline
$\mathbf{y}_l, l\in\mathcal{L}$ & Signals received by SUs\\
\hline
$\mathbf{y}_k, k\in\mathcal{K}$ & Signals received by PUs\\
\hline
$\mathbf{H}_l, l\in\mathcal{L}$ & Channel between the SBS and SUs\\
\hline
${\mathbf{n}}_l, l\in\mathcal{L}$ & Equivalent noise at the SUs\\
\hline
${\mathbf{n}}_k, k\in\mathcal{K}$ & Noise at the PUs\\
\hline
$\mathbf{D}_l, l\in\mathcal{L}$ & Channels between the SBS and SUs\\
\hline
$\mathbf{D}_k, k\in\mathcal{K}$ & Channels between the PBS and PUs\\
\hline
$\hat{\mathbf{D}}_l, l\in\mathcal{L}$ & Channels between the PBS and SUs\\
\hline
$\hat{\mathbf{D}}_k, k\in\mathcal{K}$ & Channels between the SBS and PUs\\
\hline
${\mathbf{G}}_l, l\in\mathcal{L}$ & Channels from the STAR-RIS to SUs\\
\hline
${\mathbf{G}}_k, k\in\mathcal{K}$ & Channels from the STAR-RIS to PUs\\
\hline
${\mathbf{x}}_p$ & Signal from the PBS\\
\hline
\end{tabular}
\end{table}

\section{System Model} \label{sec:system_model}
In this section, we begin by outlining the system architecture of the STAR-RIS aided MIMO CR communication system. Subsequently, we describe the signal model and introduce the STAR-RIS models. Finally, we formulate the problem of maximizing the sum rate for the secondary users.
\subsection{System Architecture}
\begin{figure}[!htbp]
\centering
\includegraphics[width=0.5\textwidth]{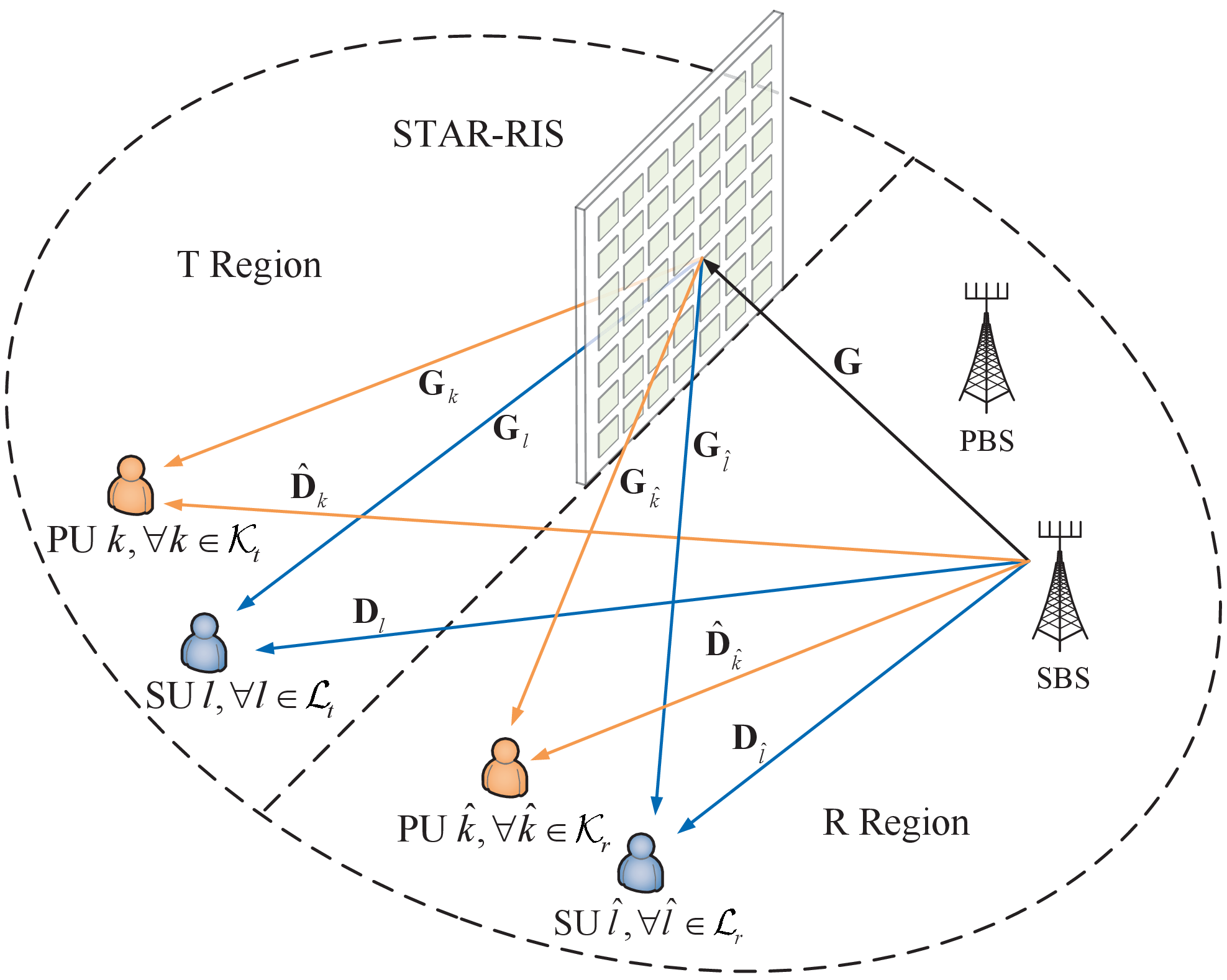}
\caption{{The proposed STAR-RIS aided cognitive radio network.}}
\label{system_model_multiple}
\end{figure}
{As shown in Fig.~\ref{system_model_multiple}, a STAR-RIS aided MIMO CR system is investigated in this work, where the secondary BS (SBS) communicates with $L$ secondary users (SUs) by reusing the spectrum resource occupied by the primary BS (PBS) and its associated $K$ primary users (PUs).} An $N$-element STAR-RIS is deployed to facilitate the communication for the secondary network and suppress the interference in the primary network. The PBS, PUs, SBS, and SUs are equipped with $M_p$, $U_p$, $M_s$, and $U_s$ antennas, respectively. The STAR-RIS divides the full space into the transmission region and the reflection region. All PUs and SUs are evenly distributed on the two sides of the STAR-RIS {(assuming $K$ and $L$ are even numbers for brevity). The indices of STAR-RIS elements, SUs, SUs in the transmission region of the STAR-RIS, and SUs in the reflection region of the STAR-RIS are collected in sets $\mathcal{N}$, $\mathcal{L}$, $\mathcal{L}_t$, $\mathcal{L}_r$, respectively. The indices of PUs, PUs in the transmission region of the STAR-RIS, and PUs in the reflection region of the STAR-RIS are collected in sets $\mathcal{K}$, $\mathcal{K}_t$, and $\mathcal{K}_r$, respectively. {{The distribution of users with respect to the STAR-RIS is assumed to be known by the SBS. This information can be obtained by the SBS during the channel estimation stage.}}} {Comparisons with Previous RIS CR Works are provided in Table~\ref{Comparisons4}.}

     \begin{table*}[h]

    \caption{{Comparisons with Previous RIS CR Works}}\label{Comparisons4}
    \centering
    \begin{threeparttable}
    {{\begin{tabular}{*{11}{l}}
    \toprule
         & \makecell[c]{Our paper} & \cite{9235486} & \cite{9599553} & \cite{zhang2021robust} & \cite{9183907} & \cite{10347404} & \cite{10345673} & \cite{10263782} & \cite{10093979} & \cite{wen2022star} \\
    \midrule
    \midrule
        STAR-RIS & \makecell[c]{\cmark} &  &  &  &  & \makecell[c]{\cmark} & \makecell[c]{\cmark} & \makecell[c]{\cmark} & \makecell[c]{\cmark} & \makecell[c]{\cmark} \\
    \midrule
        RIS & applicable & \makecell[c]{\cmark}  & \makecell[c]{\cmark}  & \makecell[c]{\cmark}  & \makecell[c]{\cmark}  &  &  & &  &  \\
    \midrule
        Single SU & applicable & \makecell[c]{\cmark} &  &  &  & \makecell[c]{\cmark} & \makecell[c]{\cmark} &  & \makecell[c]{\cmark} & \makecell[c]{\cmark}\\
    \midrule
        Multiple SUs & \makecell[c]{\cmark} &  & \makecell[c]{\cmark} & \makecell[c]{\cmark} & \makecell[c]{\cmark} &  &  &\makecell[c]{\cmark}  & &  \\
    \midrule
        SISO & applicable & & & & & \makecell[c]{\cmark} & \makecell[c]{\cmark} & \makecell[c]{\cmark} & \makecell[c]{\cmark} &  \\
    \midrule
        MISO & applicable & \makecell[c]{\cmark}& &  & \makecell[c]{\cmark} &  & & & &\makecell[c]{\cmark}  \\
    \midrule
        MIMO & \makecell[c]{\cmark} & &\makecell[c]{\cmark} & \makecell[c]{\cmark} &  &  & & &  \\
    \midrule
        Beamforming design & \makecell[c]{\cmark} &\makecell[c]{\cmark} & \makecell[c]{\cmark}& \makecell[c]{\cmark}&\makecell[c]{\cmark} &  &  &  &&  \makecell[c]{\cmark}\\
    \midrule
        Performance analysis &  & & & & & \makecell[c]{\cmark} & \makecell[c]{\cmark} & \makecell[c]{\cmark} & \makecell[c]{\cmark} &\\
    \bottomrule
    \end{tabular}}}
    \begin{tablenotes}
        \item{{SISO and MISO mean single-input single-output and multiple-input single-output, respectively. For STAR-RIS aided CR works~\cite{10347404,10345673,10263782,10093979,wen2022star}, Single SU and Multiple SUs refer to single SU on both sides of the STAR-RIS and multiple SUs on both sides of the STAR-RIS, respectively.}}
    \end{tablenotes}
    \end{threeparttable}
    \end{table*}

\subsection{STAR Model}
{Let $\mathbf{\Phi}_{i}=\diag\left(\sqrt{\rho_{1}^i}e^{j\theta_{1}^i}, \sqrt{\rho_{2}^i}e^{j\theta_{2}^i}, \cdots, \sqrt{\rho_{N}^i}e^{j\theta_{N}^i}\right), \forall i\in\{t,r\}$ represent the transmission and reflection coefficient matrix of the STAR-RIS}, where $\rho_n^t, \rho_n^r$ are the amplitude coefficients of the $n$-th STAR element and $\theta_n^t, \theta_n^r$ are the phase-shifts of the $n$-th STAR element. The STAR-RIS works under ES mode with all elements transmitting and reflecting signals at the same time.

We first introduce the {independent phase-shift model} where the transmission and reflection phases of the same STAR element can be adjusted independently and the transmission and reflection amplitudes of the same STAR element are constrained by the law of conservation of energy~\cite{9690478}. The coefficients for the $n$-th STAR element should satisfy:
\begin{equation}\label{independent}
\begin{aligned}
\mathcal{C}=\left\{ {\begin{array}{*{20}{c}}
{\rho _n^t,\rho _n^r}&\vline& {\rho _n^t,\rho _n^r \in [0,1],\rho _n^t + \rho _n^r = 1}\\
{\theta _n^t,\theta _n^r}&\vline& {\theta _n^t,\theta _n^r \in [0,2\pi ), \forall n\in\mathcal{N}}
\end{array}} \right\}.
\end{aligned}
\end{equation}
To achieve STAR-RIS with independent transmission and reflection phases for each element, the STAR elements must either be non-passive or lossy, leading to potential drawbacks such as high power consumption or degradation in system performance. {For passive and lossless STAR-RISs, the transmitted and reflected phases of each of its elements must adhere to the following constraints~\cite{9774942}:}
\begin{equation}\label{coupled}
\begin{aligned}
\widehat{\mathcal{C}}=\left\{ {\begin{array}{*{20}{c}}
{\rho _n^t,\rho _n^r}&\vline& {\rho _n^t,\rho _n^r \in [0,1],\rho _n^t + \rho _n^r = 1, \forall n\in\mathcal{N}}\\
{\theta _n^t,\theta _n^r}&\vline& {\cos(\theta _n^t-\theta _n^r)=0, \theta _n^t,\theta _n^r \in [0,2\pi )}
\end{array}} \right\}.
\end{aligned}
\end{equation}
We termed this STAR-RIS model as the {coupled phase-shift model}. To comprehensively explore the role of the STAR-RIS in the proposed system, we consider both independent and coupled STAR-RIS models in this work.
\subsection{Signal Model}
The signal sent by the SBS can be expressed as:
\begin{equation}
\begin{aligned}
{\mathbf{x}=\sum_{l=1}^{L}\mathbf{w}_ls_l,}
\end{aligned}
\end{equation}
{where $\mathbf{s}_l$ and $\mathbf{w}_l\in\mathbb{C}^{M_s \times 1}$ represent the data symbol and the SBS beamformer for the SU $l$, respectively. The received signals at the SUs and  PUs, denoted as $\mathbf{y}_l, \forall l \in \mathcal{L}_{i}, i \in \{t,r\}$} and $\mathbf{y}_k, \forall k \in \mathcal{K}_{i}, i \in \{t,r\}$, are given by
\begin{equation}
\begin{aligned}
\!\!\!{\mathbf{y}_l}&{=(\mathbf{D}_{l}+\mathbf{G}_{l}\mathbf{\Phi}_{i}\mathbf{G}_s)\mathbf{x}+(\hat{\mathbf{D}}_{l}+\mathbf{G}_{l}\mathbf{\Phi}_{i}\mathbf{G}_p)\mathbf{x}_p+\mathbf{n}_l}\\
&{=(\mathbf{D}_{l}+\mathbf{G}_{l}\mathbf{\Phi}_{i}\mathbf{G}_s)\mathbf{w}_ls_l+(\mathbf{D}_{l}+\mathbf{G}_{l}\mathbf{\Phi}_{i}\mathbf{G}_s)\sum_{j\ne l}\mathbf{w}_{j}s_{j}+\hat{\mathbf{n}}_l}\\
&{=\mathbf{H}_{l}\mathbf{w}_ls_l+\sum_{j\ne l}\mathbf{H}_{l}\mathbf{w}_js_j+\hat{\mathbf{n}}_l,}
\end{aligned}
\end{equation}
\begin{equation}
\begin{aligned}
\mathbf{y}_k=(\mathbf{D}_{k}+\mathbf{G}_{k}\mathbf{\Phi}_{i}\mathbf{G}_p)\mathbf{x}_p+(\hat{\mathbf{D}}_{k}+\mathbf{G}_{k}\mathbf{\Phi}_{i}\mathbf{G}_s)\mathbf{x}+\mathbf{n}_k,
\end{aligned}
\end{equation}
where channels from the SBS to the SUs, from the SBS to the PUs, from the SBS to the STAR-RIS, from the PBS to the SUs, from the PBS to the PUs, from the PBS to the STAR-RIS, from the STAR-RIS to the SUs, and from the STAR-RIS to the PUs are represented with matrices ${\mathbf{D}_l\in\mathbb{C}^{U_s\times M_s}}$, $\hat{\mathbf{D}}_k\in\mathbb{C}^{U_p\times M_s}$, $\mathbf{G}_s\in\mathbb{C}^{N\times M_s}$, ${\hat{\mathbf{D}}_l\in\mathbb{C}^{U_s\times M_p}}$, $\mathbf{D}_k\in\mathbb{C}^{U_p\times M_p}$, $\mathbf{G}_p\in\mathbb{C}^{N\times M_p}$, ${\mathbf{G}_l\in\mathbb{C}^{U_s\times N}}$, and $\mathbf{G}_k\in\mathbb{C}^{U_p\times N}$, respectively. Matrix ${\mathbf{H}_l=\mathbf{D}_{l}+\mathbf{G}_{l}\mathbf{\Phi}_{i}\mathbf{G}_s}$ is the aggregation channel from the SBS to SUs. {The quasi-static flat-fading model is adopted for all channels. The channels are assumed to remain the same within each time slot and change across different time slots. During the channel estimation phase, the channels among the SBS, SUs, PUs, and the STAR-RIS can be obtained with existing methods~\cite{wu2021channel,guo2022uplink}.} Vector $\mathbf{x}_p\in\mathbb{C}^{M_p\times 1}$ denotes the transmit signal from the PBS. The beamforming design is not considered at the PBS in this work. {The reason can be summarized as follows: Equation $(4)$ demonstrates that the beamforming design at the PBS affects the interference power encountered by the SUs.  However, optimizing the beamformers at the PBS to reduce interference at the SUs requires additional CSI between the PBS and SUs, which introduces significant channel estimation overheads and is challenging to achieve in CR systems. To address this challenge, the interference from the PBS to the SUs is approximated using additive white Gaussian noise terms~\cite{9183907}, and the exact structure of PBS beamformers is not considered in this work.} \noindent ${\hat{\mathbf{n}}_l=(\hat{\mathbf{D}}_{l}+\mathbf{G}_{l}\mathbf{\Phi}_{i}\mathbf{G}_p)\mathbf{x}_p+\mathbf{n}_l\sim\mathcal{CN}(\mathbf{0},\sigma_l^2\mathbf{I})}$ is the equivalent noise at the SUs, where ${\sigma_l^2=\hat{\sigma}_l^2+\sigma_s^2}$. ${\hat{\sigma}_l^2}$ accounts for the interference from PBS and $\sigma_s^2$ accounts for the thermal noise at the SUs. {$\mathbf{n}_k\sim\mathcal{CN}(\mathbf{0},\sigma_p^2\mathbf{I})$ stands for the additive white Gaussian noise at the $k$-th PU}.

{The received SINR of the SU $l$ can be written as}
\begin{equation}
\begin{aligned}
&\text{SINR}_l=\\
&{\mathbf{w}_l^H\mathbf{H}_l^H\left(\sum_{j\ne l}\mathbf{H}_l\mathbf{w}_{j}\mathbf{w}_{j}^H\mathbf{H}_l^H+\sigma_l^2\mathbf{I}\right)^{-1}\mathbf{H}_l\mathbf{w}_l, \forall l \in \mathcal{L}.}
\end{aligned}
\end{equation}

The interference imposed by the SBS at the $k$-th PU, {referred to as the interference temperature (IT) in CR systems}, can be expressed as
\begin{equation}
\begin{aligned}
&\text{IT}_k=\\
&{\quad\sum_{l\in\mathcal{L}}\|(\hat{\mathbf{D}}_{k}+\mathbf{G}_{k}\mathbf{\Phi}_{i}\mathbf{G}_s)\mathbf{w}_l\|_2^2, \forall k \in \mathcal{K}_i, \forall i \in \{t,r\}.}
\end{aligned}
\end{equation}
{The IT thresholds are related to STAR-RIS coefficients considering that the STAR-RIS is shared by both the primary and second systems. The change of STAR-RIS coefficients has an influence on the IT at PUs.}
\subsection{Optimization Problem Formulation}
The SBS beamformers ${\mathbf{w}\buildrel \Delta \over = \{\mathbf{w}_1, \mathbf{w}_2, \cdots, \mathbf{w}_L\}}$ and the transmission- and reflection-coefficient matrices $\mathbf{\Phi}\buildrel \Delta \over =\{\mathbf{\Phi}_{t},\mathbf{\Phi}_{r}\}$ are optimized to maximize the sum rate of the secondary network subject to the constraints introduced by the SBS, PUs, and the STAR-RIS.
\begin{subequations}\label{problem:sum_rate}
    \begin{align}        
        \max_{\mathbf{w},\mathbf{\Phi}} \quad &  {\sum_{l\in\mathcal{L}}\log\left(1+\text{SINR}_l\right)} \\
        \label{constraint:STAR}
        \mathrm{s.t.} \quad & \mathbf{\Phi}_i\in \mathcal{F}_j, \forall i \in\{t,r\}, \forall j \in\{1,2\},\\
        \label{constraint:power}
        & {\sum_{l\in\mathcal{L}}\|\mathbf{w}_l\|^{2}\le P_s,}\\
        \label{constraint:IT}
        & \text{IT}_k\le \Gamma_k, \forall k \in \mathcal{K},
    \end{align}
\end{subequations}
where the constraint~\eqref{constraint:STAR} is the ES coefficients constraint for the STAR-RIS. When the \emph{independent phase-shift model} is used, the STAR coefficients are constrained by~\eqref{independent}, with $\mathcal{F}_1$ characterizing the corresponding feasible set for the STAR matrices. If the {coupled phase-shift model} is employed, the STAR coefficients are constrained by~\eqref{coupled}, with $\mathcal{F}_2$ characterizing the corresponding feasible set for the STAR matrices. The constraint~\eqref{constraint:power} is the power constraint of the SBS, with $P_s$ denoting the power budget of the SBS. The constraint~\eqref{constraint:IT} guarantees that the maximum interference leakage at the PUs is tolerable with $\Gamma_k$ representing the IT threshold of the $k$-th PU. 

The formulated optimization problem can be extended to the MS STAR-RIS case by adding the constraint $\rho _n^t,\rho _n^r \in \{0,1\}, \forall n \in\mathcal{N}$. A penalty method can be adopted to handle this constraint. Due to page limitations, we do not provide a detailed derivation of the penalty method and refer readers to works~\cite{ben1997penalty,mu2021simultaneously}.
\section{Joint Beamforming Algorithm with Independent STAR-RIS Model} 
This section proposes a BCD-based method to solve \eqref{problem:sum_rate} for STAR-RIS with phase shifts that are independent of each other.
\subsection{Problem Reformulation}
The challenge in optimizing $\mathbf{w}$ and $\mathbf{\Phi}$ arises from the sum of logarithmic functions in the objective function in~\eqref{problem:sum_rate}. To tackle this difficulty, the Lagrangian dual transform is adopted to reformulate the optimization problem stated in~\eqref{problem:sum_rate}~\cite{shen2018fractional1}. Specifically, by introducing auxiliary variables ${\bm{\gamma}=[\gamma_1, \gamma_2, \cdots, \gamma_L]^T}$, problem~\eqref{problem:sum_rate} can be equivalently written as
\begin{subequations}\label{problem:sum_rate_1}
    \begin{align}        
        \max_{\mathbf{w},\mathbf{\Phi},\bm{\gamma}}  \  &  {\sum_{l\in\mathcal{L}}\log\left(1+\gamma_l\right)} \\   
        \mathrm{s.t.} \quad & \eqref{constraint:STAR}-\eqref{constraint:IT}, \\
        \label{gamma}
        & {\gamma_l \le \mathbf{w}_l^H\mathbf{H}_l^H\left(\sum_{j\ne l}\mathbf{H}_l\mathbf{w}_{j}\mathbf{w}_{j}^H\mathbf{H}_l^H+\sigma_l^2\mathbf{I}\right)^{-1}\!\!\!\mathbf{H}_l\mathbf{w}_l, \forall l}.
    \end{align}
\end{subequations}
This optimization problem can be sovled sequtially over $\bm{\gamma}$ and $\{\mathbf{w},\mathbf{\Phi}\}$. First, with fixed $\{\mathbf{w},\mathbf{\Phi}\}$, we have following optimization problem
\begin{subequations}\label{problem:sum_rate_gamma}
    \begin{align}        
        \max_{\bm{\gamma}}  \quad &  {\sum_{l\in\mathcal{L}}\log\left(1+\gamma_l\right)} \\   
        \mathrm{s.t.} \quad & \eqref{gamma}.
    \end{align}
\end{subequations}
Problem~\eqref{problem:sum_rate_gamma} is convex with respect to $\bm{\gamma}$ and thus the strong  duality holds~\cite{boyd2004convex}. And the optimal solution of $\gamma_l, \forall l,$ can be solved as
\begin{equation}\label{gammaopt}
    {\gamma_l^*=\mathbf{w}_l^H\mathbf{H}_l^H\left(\sum_{j\ne l}\mathbf{H}_l\mathbf{w}_{j}\mathbf{w}_{j}^H\mathbf{H}_l^H+\sigma_l^2\mathbf{I}\right)^{-1}\mathbf{H}_l\mathbf{w}_l.}
\end{equation}
The Lagrangian function of problem~\eqref{problem:sum_rate_gamma} can be expressed as in~\eqref{Lagrangian} at the top of the next page, where ${\bm{\lambda}=[\lambda_1, \lambda_1, \cdots, \lambda_L]^T}$ is the non-negative dual variables for constraint~\eqref{gamma}. 
\begin{figure*}[!t]
\normalsize
\begin{equation}\label{Lagrangian}
{L(\bm{\gamma},\bm{\lambda})=\sum_{l\in\mathcal{L}}\log\left(1+\gamma_l\right)-\sum_{l\in\mathcal{L}}\lambda_l\left(\gamma_l-\mathbf{w}_l^H\mathbf{H}_l^H\left(\sum_{j\ne l}\mathbf{H}_l\mathbf{w}_{j}\mathbf{w}_{j}^H\mathbf{H}_l^H+\sigma_l^2\mathbf{I}\right)^{-1}\mathbf{H}_l\mathbf{w}_l\right).}
\end{equation}
\hrulefill \vspace*{0pt}
\end{figure*}
The strong duality holds for problem~\eqref{problem:sum_rate_gamma} and we have
\begin{equation}\label{strong_duality}
    \max_{\bm{\gamma}}\min_{\bm{\lambda}\succeq 0} \quad L(\bm{\gamma},\bm{\lambda})=\min_{\bm{\lambda}\succeq 0}\max_{\bm{\gamma}} \quad L(\bm{\gamma},\bm{\lambda}).
\end{equation}
The left hand side of~\eqref{strong_duality} is equivalent to problem~\eqref{problem:sum_rate_gamma}. The right hand side of~\eqref{strong_duality} is the dual problem of~\eqref{problem:sum_rate_gamma}, which can also be given as
\begin{subequations}\label{problem:sum_rate_dual}
    \begin{align}
        \min_{\bm{\lambda}} \quad & L(\bm{\gamma}^*,\bm{\lambda})\\
        \mathrm{s.t.} \quad & \bm{\lambda}  \succeq 0.
    \end{align}
\end{subequations}
Problem~\eqref{problem:sum_rate_dual} can be optimally solved as:
\begin{equation}\label{lambdaopt}
\begin{aligned}
    \lambda^*_l=&{\frac{1}{1+\gamma_l^*}}\\
    =&{\frac{1}{1+\mathbf{w}_l^H\mathbf{H}_l^H\left(\sum_{j\ne l}\mathbf{H}_l\mathbf{w}_{j}\mathbf{w}_{j}^H\mathbf{H}_i^H+\sigma_l^2\mathbf{I}\right)^{-1}\mathbf{H}_l\mathbf{w}_l},}
\end{aligned}
\end{equation}
Substituting~\eqref{lambdaopt} into the left hand side of~\eqref{strong_duality}, we can rewrite the problem~\eqref{problem:sum_rate_gamma} as
\begin{subequations}
    \begin{align}
        \max_{\bm{\gamma}} \quad & L(\bm{\gamma},\bm{\lambda}^*).
    \end{align}
\end{subequations}
Furthermore, considering the optimization over $\{\mathbf{w},\mathbf{\Phi}\}$, problem~\eqref{problem:sum_rate_1} is reformulated as follows:
\begin{subequations}\label{problem:sum_rate_LDR}
    \begin{align}   
    \label{obj:sum_rate_LDR}     
        \max_{\bm{\gamma},\mathbf{w},\mathbf{\Phi},} \quad & f_{\text{LDT}}\left(\bm{\gamma},\mathbf{w},\mathbf{\Phi}\right) \\
        \mathrm{s.t.} \quad & \eqref{constraint:STAR}-\eqref{constraint:IT},
    \end{align}
\end{subequations}
where the objective function of~\eqref{problem:sum_rate_LDR} can be expressed as in~\eqref{LDR} at the top of next page,
\begin{figure*}[!t]
\normalsize
\begin{equation}\label{LDR}
\begin{aligned}
f_{\text{LDT}}\left(\bm{\gamma},\mathbf{w},\mathbf{\Phi}\right)=&{\sum_{l\in\mathcal{L}}\log\left(1+\gamma_l\right)-\sum_{l\in\mathcal{L}}\frac{\gamma_l-\mathbf{w}_l^H\mathbf{H}_l^H\left(\sum_{j\ne l}\mathbf{H}_l\mathbf{w}_{j}\mathbf{w}_{j}^H\mathbf{H}_l^H+\sigma_l^2\mathbf{I}\right)^{-1}\mathbf{H}_l\mathbf{w}_l}{1+\mathbf{w}_l^H\mathbf{H}_l^H\left(\sum_{j\ne l}\mathbf{H}_l\mathbf{w}_{j}\mathbf{w}_{j}^H\mathbf{H}_l^H+\sigma_l^2\mathbf{I}\right)^{-1}\mathbf{H}_l\mathbf{w}_l}}\\
=&{\sum_{l\in\mathcal{L}}\log\left(1+\gamma_l\right)-\sum_{l\in\mathcal{L}}\Bigg(\frac{\left(1+\gamma_l\right)}{1+\mathbf{w}_l^H\mathbf{H}_l^H\left(\sum_{j\ne l}\mathbf{H}_l\mathbf{w}_{j}\mathbf{w}_{j}^H\mathbf{H}_l^H+\sigma_l^2\mathbf{I}\right)^{-1}\mathbf{H}_l\mathbf{w}_l}-1\Bigg)}\\
=&{\sum_{l\in\mathcal{L}}\log\left(1+\gamma_l\right)-\sum_{l\in\mathcal{L}}\gamma_l+\sum_{l\in\mathcal{L}}\left(1+\gamma_l\right)\mathbf{w}_l^H\mathbf{H}_l^H\mathbf{J}_l^{-1}\mathbf{H}_l\mathbf{w}_l.}
\end{aligned}
\end{equation}
\end{figure*}
 wherein  
\begin{equation}
\begin{aligned}
{\mathbf{J}_l=\sum_{j\in\mathcal{L}}\mathbf{H}_l\mathbf{w}_{j}\mathbf{w}_{j}^H\mathbf{H}_l^H+\sigma_l^2\mathbf{I},}
\end{aligned}
\end{equation}
and the last equality in~\eqref{LDR} follows from the Woodbury matrix identity, which can be expressed as in~\eqref{woodbury} at the top of the next page.
\begin{figure*}[!t]
\normalsize
\begin{equation}\label{woodbury}
\begin{aligned}
{\Bigg(1+\mathbf{w}_l^H\mathbf{H}_l^H\Big(\sum_{j\ne l}\mathbf{H}_l\mathbf{w}_{j}\mathbf{w}_{j}^H\mathbf{H}_l^H+\sigma_l^2\mathbf{I}\Big)^{-1}\!\!\mathbf{H}_l\mathbf{w}_l\Bigg)^{-1}\!\!}&{=1-\mathbf{w}_l^H\mathbf{H}_l^H\Big(\mathbf{H}_l\mathbf{w}_{l}\mathbf{w}_{l}^H\mathbf{H}_l^H+\sum_{j\ne l}\mathbf{H}_l\mathbf{w}_{j}\mathbf{w}_{j}^H\mathbf{H}_l^H+\sigma_l^2\mathbf{I}\Big)^{-1}\!\!\mathbf{H}_l\mathbf{w}_l}\\
&{=1-\mathbf{w}_l^H\mathbf{H}_l^H\mathbf{J}_l^{-1}\mathbf{H}_l\mathbf{w}_l.}
\end{aligned}
\end{equation}
\hrulefill \vspace*{0pt}
\end{figure*}
The problem~\eqref{problem:sum_rate_LDR} can represent the reformulated sum rate maximization problem with {independent phase-shift model} and {coupled phase-shift model} when the constraint~\eqref{constraint:STAR} takes a different form.

Then, we concentrate on the {independent phase-shift model} and use the BCD approach to optimize $\mathbf{\Phi}$, $\bm{\gamma}$, and $\mathbf{w}$ in turn.
\subsection{Optimizing $\bm{\gamma}$ for Given $\mathbf{w}$ and $\mathbf{\Phi}$}
With fixed SBS beamformers $\mathbf{w}$ and STAR-RIS coefficients $\mathbf{\Phi}$, the sub-problem for $\bm{\gamma}$ is expressed as follows:
\begin{equation}
    \max_{\bm{\gamma}} \quad f_{\text{LDT}}\left(\bm{\gamma},{\mathbf{w}},{\mathbf{\Phi}}\right).
\end{equation}
{The optimal $\gamma_l$ can be obtained by solving $\partial L/\partial {\gamma _l} = 0$. Specifically, the optimal $\gamma_l$ is given by}
\begin{equation}\label{gamma_opt}
\begin{aligned}
{\gamma_l^{\text{opt}}=\frac{\mathbf{w}_l^H\mathbf{H}_l^H\mathbf{J}_l^{-1}\mathbf{H}_l\mathbf{w}_l}{1-\mathbf{w}_l^H\mathbf{H}_l^H\mathbf{J}_l^{-1}\mathbf{H}_l\mathbf{w}_l}, \forall l \in\mathcal{L}.}
\end{aligned}
\end{equation}

\subsection{Optimizing $\mathbf{w}$ for Given $\bm{\gamma}$ and $\mathbf{\Phi}$}
For given $\bm{\gamma}$ and $\mathbf{\Phi}$, the subproblem with respect to $\mathbf{w}$ is expressed as:
\begin{subequations}\label{problem:sum_rate_transmit_beamforming}
    \begin{align}        
         \max_{\mathbf{w}} \quad & {\sum_{l\in\mathcal{L}}\left(1+\gamma_l\right)\mathbf{w}_l^H\mathbf{H}_l^H\mathbf{J}_l^{-1}\mathbf{H}_l\mathbf{w}_l} \\
        \mathrm{s.t.} \quad & \eqref{constraint:power}, \eqref{constraint:IT}.
    \end{align}
\end{subequations}
 To tackle the non-convex multidimensional multiple-ratio fractional programming problem~\eqref{problem:sum_rate_transmit_beamforming}, we use the CQT~\cite{8310563} to reformulate problem~\eqref{problem:sum_rate_transmit_beamforming} as follows:
\begin{subequations}\label{problem:sum_rate_transmit_beamforming_CQT}
    \begin{align}        
         \max_{\mathbf{w},\bm{\alpha}} \quad & f_{\text{CQT}}\left(\mathbf{w},\bm{\alpha}\right) \\
        \mathrm{s.t.} \quad & \eqref{constraint:power}, \eqref{constraint:IT},
    \end{align}
\end{subequations}
where ${\bm{\alpha}\buildrel \Delta \over =\{\bm{\alpha}_1, \bm{\alpha}_2, \cdots, \bm{\alpha}_L\}}$ are auxiliary variables. The objective function of~\eqref{problem:sum_rate_transmit_beamforming_CQT} can be expressed as
\begin{equation}\label{CQT}
\begin{aligned}
f_{\text{CQT}}&\left(\mathbf{w},\bm{\alpha}\right)=\\
&{\sum_{l\in\mathcal{L}}(2\sqrt{\left(1+\gamma_l\right)}\Re(\bm{\alpha}_l^H\mathbf{H}_l\mathbf{w}_l)-\bm{\alpha}_l^H\mathbf{J}_l\bm{\alpha}_l).}
\end{aligned}
\end{equation}

\subsubsection{Optimizing $\bm{\alpha}$ for given $\mathbf{w}$} {With fixed BS beamforming vectors $\mathbf{w}$, the optimization problem with respect to $\bm{\alpha}$ can be given as} 
\begin{equation}
    {\max_{\bm{\alpha}} \quad  f_{\text{CQT}}\left(\mathbf{w},\bm{\alpha}\right).}
\end{equation}
{The optimal $\bm{\alpha}_l$ can be obtained by solving $\partial L/\partial {\bm{\alpha} _l} = 0$. Specifically, the optimal $\bm{\alpha}_l$ is given by}
\begin{equation}\label{alpha}
\begin{aligned}
{\bm{\alpha}_l^{\text{opt}}=\sqrt{\left(1+\gamma_l\right)}\mathbf{J}_l^{-1}\mathbf{H}_l\mathbf{w}_l, \forall l\in\mathcal{L}.}
\end{aligned}
\end{equation}
Substituting the solved $\bm{\alpha}_l^{\text{opt}}, \forall l\in\mathcal{L}$ into the objective function of problem~\eqref{problem:sum_rate_transmit_beamforming_CQT} leads to the objective function of problem~\eqref{problem:sum_rate_transmit_beamforming}. The equivalence between problem~\eqref{problem:sum_rate_transmit_beamforming} and problem~\eqref{problem:sum_rate_transmit_beamforming_CQT} is therefore proved.
\subsubsection{Optimizing  $\mathbf{w}$ for given $\bm{\alpha}$ }  
To facilitate the following derivation, we define
\begin{equation}
\begin{aligned}
{\mathbf{A}=\sum_{l\in\mathcal{L}}\mathbf{H}_l^H\bm{\alpha}_l\bm{\alpha}_l^H\mathbf{H}_l,}
\end{aligned}
\end{equation}
\begin{equation}
\begin{aligned}
\mathbf{A}_k=\left(\hat{\mathbf{D}}_k+\mathbf{G}_k\mathbf{\Phi}_i\mathbf{G}_s\right), \forall k \in \mathcal{K}_i, \forall i \in \{t,r\},
\end{aligned}
\end{equation}
\begin{equation}
\begin{aligned}
{\bm{a}_l^H=\sqrt{1+\gamma_l}\bm{\alpha}_l^H\mathbf{H}_l, \forall l \in \mathcal{L}.
}\end{aligned}
\end{equation}
With fixed auxiliary variables $\bm{\alpha}$, the BS beamforming optimization problem can be given as follows:
\begin{subequations}\label{problem:w}
    \begin{align}        
        \min_{{\mathbf{w}}} \quad & {\sum_{l\in\mathcal{L}}({\mathbf{w}_l}^H\mathbf{A}{\mathbf{w}_l}-2\Re(\bm{a}_l^H{\mathbf{w}_l}))}  \\
        \mathrm{s.t.} \quad & {\sum_{l\in\mathcal{L}}\|{\mathbf{w}_l}\|^{2}\le P_s,}\\
        & {\sum_{l\in\mathcal{L}}{\mathbf{w}_l}^H\mathbf{A}_k{\mathbf{w}_l} \le \Gamma_k.}
    \end{align}
\end{subequations}
{Problem~\eqref{problem:w} is a convex problem and its optimal solution can be obtained via CVX~\cite{grant2014cvx}.}
\subsection{Optimizing $\mathbf{\Phi}$ for Given $\bm{\gamma}$ and $\mathbf{w}$}
{Note that the aggregation channel from the SBS to the SU~$l$, i.e., $\mathbf{H}_l(\mathbf{\Phi}_i)=\mathbf{D}_{l}+\mathbf{G}_{l}\mathbf{\Phi}_{i}\mathbf{G}_s, \forall l\in\mathcal{L}_i, \forall i\in\{t,r\}$, are functions of STAR transmission and reflection matrices. For given $\bm{\gamma}$ and $\mathbf{w}$, the subproblem with respect to $\mathbf{\Phi}$ can be given as follows:}
\begin{subequations}\label{problem:sum_rate_STAR_beamforming}
    \begin{align}        
         \max_{\mathbf{\Phi}} \quad & {\sum_{l\in\mathcal{L}}\left(1+\gamma_l\right)\mathbf{w}_l^H\mathbf{H}_l^H(\mathbf{\Phi}_i)\mathbf{J}_l^{-1}\mathbf{H}_l(\mathbf{\Phi}_i)\mathbf{w}_l} \\
        \mathrm{s.t.} \quad & \eqref{constraint:STAR}, \eqref{constraint:IT}.
    \end{align}
\end{subequations}
Similar to problem~\eqref{problem:sum_rate_transmit_beamforming}, we use the CQT to reformulate problem~\eqref{problem:sum_rate_STAR_beamforming} as follows:
\begin{subequations}\label{problem:sum_rate_STAR_beamforming_CQT}
    \begin{align}        
         \max_{\mathbf{\Phi},\bm{\beta}} \quad & g_{\text{CQT}}\left(\mathbf{\Phi},\bm{\beta}\right) \\
        \mathrm{s.t.} \quad & \eqref{constraint:STAR}, \eqref{constraint:IT},
    \end{align}
\end{subequations}
where ${\bm{\beta}\buildrel \Delta \over =\{\bm{\beta}_1,\bm{\beta}_2,\cdots, \bm{\beta}_L\}}$ are auxiliary variables. The objective function of~\eqref{problem:sum_rate_STAR_beamforming_CQT} can be expressed as
\begin{equation}\label{CQT_STAR}
\begin{aligned}
g_{\text{CQT}}&\left(\mathbf{\Phi},\bm{\beta}\right)=\\
&{\sum_{l\in\mathcal{L}}(2\sqrt{\left(1+\gamma_l\right)}\Re(\bm{\beta}_l^H\mathbf{H}_l(\mathbf{\Phi}_i)\mathbf{w}_l)-\bm{\beta}_l^H\mathbf{J}_l\bm{\beta}_l).}
\end{aligned}
\end{equation}
\subsubsection{Optimizing $\bm{\beta}$ for given $\mathbf{\Phi}$} {With fixed STAR transmission and reflection matrices $\mathbf{\Phi}$, the optimization problem with respect to $\bm{\beta}$ can be given as}
\begin{equation}
    {\max_{\bm{\beta}} \quad g_{\text{CQT}}\left(\mathbf{\Phi},\bm{\beta}\right).}
\end{equation}
{The optimal $\bm{\beta}_l$ can be obtained by solving $\partial L/\partial {\bm{\beta} _l} = 0$. Specifically, the optimal $\bm{\beta}_l$ is given by}
\begin{equation}\label{beta}
\begin{aligned}
{\bm{\beta}_l^{\text{opt}}=\sqrt{\left(1+\gamma_l\right)}\mathbf{J}_l^{-1}\mathbf{H}_l(\mathbf{\Phi}_i)\mathbf{w}_l, \forall l\in\mathcal{L}_i, \forall i\in\{t,r\}.}
\end{aligned}
\end{equation}
Substituting the solved $\bm{\beta}_l^{\text{opt}}, \forall l\in\mathcal{L}$ into the objective function of problem~\eqref{problem:sum_rate_STAR_beamforming_CQT} leads to the objective function of problem~\eqref{problem:sum_rate_STAR_beamforming}. The equivalence between problem~\eqref{problem:sum_rate_STAR_beamforming} and problem~\eqref{problem:sum_rate_STAR_beamforming_CQT} is therefore proved.
\subsubsection{Optimizing  $\mathbf{\Phi}$ for given $\bm{\beta}$ }  
With fixed auxiliary variables $\bm{\beta}$, the STAR-RIS transmission and reflection matrices optimization problem can be given as follows:
\begin{subequations}\label{problem:phi}
    \begin{align}
        \label{obj:phi}        
        \min_{\mathbf{\Phi}} \quad&  g(\mathbf{\Phi}) \\
        \label{constraint:STAR_ind} 
        \mathrm{s.t.} \quad & \mathbf{\Phi}_i\in \mathcal{F}_1, \forall i \in\{t,r\},\\
        & \eqref{constraint:IT},
    \end{align}
\end{subequations}
where the objective function of problem~\eqref{problem:phi} is given in~\eqref{g_phi} at the top of next page.
\begin{figure*}[!t]
\normalsize
\begin{equation}\label{g_phi}
\begin{aligned}
{g(\mathbf{\Phi})=\sum_{l\in\mathcal{L}}\Big(\sum_{j\in\mathcal{L}}\big|\bm{\beta}_l^H\mathbf{H}_l(\mathbf{\Phi}_i)\mathbf{w}_j\big|^2-2\sqrt{1+\gamma_l}\Re\big(\bm{\beta}_l^H\mathbf{H}_l(\mathbf{\Phi}_i)\mathbf{w}_l\big)\Big)}
\end{aligned}
\end{equation}
\end{figure*}
To facilitate the following derivation, we define
\begin{equation}
\begin{aligned}
\mathbf{v}_i=&\text{diag}(\mathbf{\Phi}_i)\\
=&\left[\sqrt{\rho_1^i}e^{j\theta_1^i}, \sqrt{\rho_2^i}e^{j\theta_2^i}, \cdots, \sqrt{\rho_N^i}e^{j\theta_N^i}\right]^T, \forall i \in\{t,r\}.
\end{aligned}
\end{equation}
Then constraint~\eqref{constraint:STAR_ind} can be reformulated as 
\begin{equation}\label{v_1}
    \mathbf{V}_i = \mathbf{v}_i\mathbf{v}_i^H, \forall i \in \{t,r\},
\end{equation}
and 
\begin{equation}\label{v_2}
   \text{diag}(\mathbf{V}_t+\mathbf{V}_r)=\mathbf{I}_N.
\end{equation}
Term ${\bm{\beta}_l^H\mathbf{H}_l(\mathbf{\Phi}_i)\mathbf{w}_j, \forall l \in\mathcal{L}_i, \forall i\in\{t,r\}}$, can be reformulated as follows:
\begin{equation}
\begin{aligned}
    {\bm{\beta}_j^H\mathbf{H}_j(\mathbf{\Phi}_i)\mathbf{w}_j}=&{\bm{\beta}_l^H\mathbf{D}_l\mathbf{w}_j+\bm{\beta}_l^H\mathbf{G}_l\mathbf{\Phi}_i\mathbf{G}_s\mathbf{w}_j}\\
    =&{\bm{\beta}_l^H\mathbf{D}_l\mathbf{w}_j+\bm{\beta}_l^H\mathbf{G}_l\text{diag}(\mathbf{G}_s\mathbf{w}_j)\mathbf{v}_{i}}\\=&{c_{lj}+\mathbf{c}_{lj}^H\mathbf{v}_{i}, \forall j \in\mathcal{L},}
\end{aligned}
\end{equation}
where ${c_{lj}=\bm{\beta}_l^H\mathbf{D}_l\mathbf{w}_j}$ and ${\mathbf{c}_{lj}^H=\bm{\beta}_l^H\mathbf{G}_l\text{diag}(\mathbf{G}_s\mathbf{w}_j)}$. Similarly, constraint~\eqref{constraint:IT} can be reformulated as in~\eqref{RE_gamma} at the top of the next page.
\begin{figure*}[!t]
\normalsize
\begin{equation}\label{RE_gamma}
\begin{aligned}
\sum_{l\in\mathcal{L}}&\|\hat{\mathbf{D}}_k\mathbf{w}_l+\mathbf{G}_k\text{diag}(\mathbf{G}_s\mathbf{w}_l)\mathbf{v}_i\|^2_2 \\&\qquad=\sum_{l\in\mathcal{L}} (\mathbf{w}_l^H\hat{\mathbf{D}}_k^H\hat{\mathbf{D}}_k\mathbf{w}_l+\mathbf{v}_i\text{diag}(\mathbf{w}_l^H\mathbf{G}_s^H)\mathbf{G}_k^H\mathbf{G}_k\text{diag}(\mathbf{G}_s\mathbf{w}_l)\mathbf{v}_i^H+2\Re(\mathbf{w}_l^H\hat{\mathbf{D}}_k\mathbf{G}_k\text{diag}(\mathbf{G}_s\mathbf{w}_l)\mathbf{v}_i))\\&\qquad \le \Gamma_k, \forall k \in \mathcal{K}_i, \forall i \in \{t,r\},
\end{aligned}
\end{equation}
\hrulefill \vspace*{0pt}
\end{figure*}
For brevity, we define
\begin{equation}
    {\mathbf{B}_k = \sum_{l\in\mathcal{L}}\text{diag}(\mathbf{w}_l^H\mathbf{G}_s^H)\mathbf{G}_k^H\mathbf{G}_k\text{diag}(\mathbf{G}_s\mathbf{w}_l),}
\end{equation}
\begin{equation}
    {\mathbf{b}_k^H = \sum_{l\in\mathcal{L}}\mathbf{w}_l^H\hat{\mathbf{D}}_k\mathbf{G}_k\text{diag}(\mathbf{G}_s\mathbf{w}_l),}
\end{equation}
\begin{equation}
    {\hat{\Gamma}_k=\Gamma_k-\sum_{l\in\mathcal{L}}\mathbf{w}_l^H\hat{\mathbf{D}}_k^H\hat{\mathbf{D}}_k\mathbf{w}_l.}
\end{equation}
Then constraint~\eqref{RE_gamma} can be rewritten as
\begin{equation}\label{RE_gamma1}
    \mathbf{v}_i^H\mathbf{B}_k\mathbf{v}_i+2\Re(\mathbf{b}_k^H\mathbf{v}_i) \le \hat{\Gamma}_k, \forall k \in \mathcal{K}_i, \forall i \in \{t,r\}.
\end{equation}
Problem~\eqref{problem:phi} can be reformulated as follows:
\begin{subequations}\label{problem:phi1}
    \begin{align}        
        \min_{\mathbf{v},\mathbf{V}} \quad&  \sum_{i\in\{t,r\}}(\mathbf{v}_i^H\mathbf{C}_i\mathbf{v}_i - 2\Re(\mathbf{c}_i^H\mathbf{v}_i)) \\
        \mathrm{s.t.} \quad & \eqref{v_1}, \eqref{v_2}, \eqref{RE_gamma1},
    \end{align}
\end{subequations}
where $\mathbf{v}=\{\mathbf{v}_t,\mathbf{v}_r\}$, $\mathbf{V}=\{\mathbf{V}_t,\mathbf{V}_r\}$, ${\mathbf{C}_i=\sum_{l\in\mathcal{L}_i}\sum_{j\in\mathcal{L}}\mathbf{c}_{lj}\mathbf{c}_{lj}^H}$, and ${\mathbf{c}_i=\sum_{l\in\mathcal{L}_i}\sqrt{1+\gamma_l}\mathbf{c}_{ll}-\sum_{l\in\mathcal{L}_i}\sum_{j\in\mathcal{L}}{c}_{ij}\mathbf{c}_{ij}}$. Problem~\eqref{problem:phi1} is non-convex due to the non-convex constraint~\eqref{v_1}, which can be recast as follows using the~\cite[Lemma~1]{6698281}:
\begin{equation}\label{v_3}
    \left[ {\begin{array}{*{20}{c}}
{{\mathbf{\Omega} _{i,1}}}&{{\mathbf{V}_i}}&{{\mathbf{v}_i}}\\
{\mathbf{V}_i^H}&{{\mathbf{\Omega} _{i,2}}}&{{\mathbf{v}_i}}\\
{\mathbf{v}_i^H}&{\mathbf{v}_i^H}&1
\end{array}} \right]\succeq 0, \forall i \in \{t,r\},
\end{equation}
\begin{equation}\label{v_4}
    \text{Tr}({\mathbf{\Omega} _{i,1}}-\mathbf{v}_i\mathbf{v}_i^H) \le 0, \forall i \in \{t,r\},
\end{equation}
where $\mathbf{\Omega} _{i,1}\in \mathbb{H}^{N \times N}$ and $\mathbf{\Omega} _{i,2}\in \mathbb{H}^{N \times N}, \forall i \in \{t,r\}$ are auxiliary matrices. Then, the SCA technique is adopted to deal with non-convex constraint~\eqref{v_4}. {In the $n$-th SCA iteration, for given point $\mathbf{v}^{(n)}=\left\{\mathbf{v}_t^{(n)},\mathbf{v}_r^{(n)}\right\}$, the global lower bound of $\text{Tr}\big({\mathbf{v}_i^{(n)}}({\mathbf{v}_i^{(n)}})^H\big), \forall i \in \{t,r\},$ can be given using the first-order Taylor expansion as follows:
\begin{equation}
\begin{aligned}\label{v_SCA}
    \text{Tr}\big(\mathbf{v}_i\mathbf{v}_i^H\big)=& \mathbf{v}_i^H\mathbf{v}_i\\
     \ge & 2\Re\Big(\big(\mathbf{v}_i^{(n)}\big)^H\mathbf{v}_i\Big)-\big(\mathbf{v}_i^{(n)}\big)^H\mathbf{v}_i^{(n)}.
    \end{aligned}
\end{equation}
This bound is tight at the given point $\mathbf{v}_i^{(n)}, \forall i \in \{t,r\}$.  Using this bound, the non-convex constraint~\eqref{v_4} can be approximated as 
\begin{equation}\label{phi_SCA}
    \text{Tr}({\mathbf{\Omega} _{i,1}}) + \big(\mathbf{v}_i^{(n)}\big)^H\mathbf{v}_i^{(n)} - 2\Re\Big(\big(\mathbf{v}_i^{(n)}\big)^H\mathbf{v}_i\Big) \le 0.
\end{equation}
With any point $\mathbf{v}_i^{(n)}, \forall i \in \{t,r\}$, and the approximation in~\eqref{phi_SCA}, problem~\eqref{problem:phi1} can be approximated as following question
\begin{subequations}\label{problem:phi_SCA}
    \begin{align}        
        \min_{\mathbf{v},\mathbf{V},\mathbf{\Omega} } \quad & \sum_{i\in\{t,r\}}(\mathbf{v}_i^H\mathbf{C}_i\mathbf{v}_i - 2\Re(\mathbf{c}_i^H\mathbf{v}_i)) \\
        \mathrm{s.t.} \quad & \eqref{v_2}, \eqref{RE_gamma1}, \eqref{v_3}, \eqref{phi_SCA},
    \end{align}
\end{subequations}
where $\mathbf{\Omega}=\{\mathbf{\Omega}_{t,1},\mathbf{\Omega}_{t,2},\mathbf{\Omega}_{r,1},\mathbf{\Omega}_{r,2}\}$. Convex problems like problem~\eqref{problem:phi_SCA} have optimal solution that can be found using CVX~\cite{grant2014cvx}. The constraint~\eqref{phi_SCA} is tighter compared to the original constraint~\eqref{v_4}. Therefore, any feasible solution to problem~\eqref{problem:phi_SCA} also serves as the feasible solution to problem~\eqref{problem:phi1}. However, a feasible solution to problem~\eqref{problem:phi1} is not necessary a feasible solution to problem~\eqref{problem:phi_SCA}. Optimal objective function of the approximation problem~\eqref{problem:phi_SCA} serves as an upper bound of that of the original problem~\eqref{problem:phi1}.

\textbf{Algorithm~\ref{algorithm1}} is a summary of the SCA algorithm for addressing subproblem~\eqref{problem:phi}. The convergence of the Algorithm~\ref{algorithm1} is discussed in the
following theorem.

\begin{theorem}\label{proposition11}
\emph{With the SCA method, the sequence generated by \text{Algorithm~\ref{algorithm1}}, i.e., $\left\{\mathbf{v}^{(n)}, n=0,1,2,\cdots\right\}$, converge to a stationary point of~\eqref{problem:phi}.} 
\begin{proof}
See Appendix A. 
\end{proof}
\end{theorem}
}

\begin{algorithm}[t]
\caption{SCA algorithm for solving problem~\eqref{problem:phi}.}\label{algorithm1}
\begin{algorithmic}[1]
\STATE {Given $\mathbf{v}^{(0)}$, $\epsilon_{\ref{algorithm1},\text{SCA}} > 0$, and set $n=0$.}
\STATE {\bf repeat:}
\STATE \quad Obtain $\hat{\mathbf{v}}^{(n)}$, by solving problem~\eqref{problem:phi_SCA} with given $\mathbf{v}^{(n)}$.
\STATE \quad Set $\mathbf{v}^{(n+1)}=\hat{\mathbf{v}}^{(n)}$, $n=n+1$.
\STATE {\bf until} the fractional increase of~\eqref{obj:phi} is below~$\epsilon_{\ref{algorithm1},\text{SCA}}$.
\end{algorithmic}
\end{algorithm}
\subsection{The Overall Algorithm}
    \begin{figure*}[!htbp]
    \centering
    \includegraphics[width=0.95\textwidth]{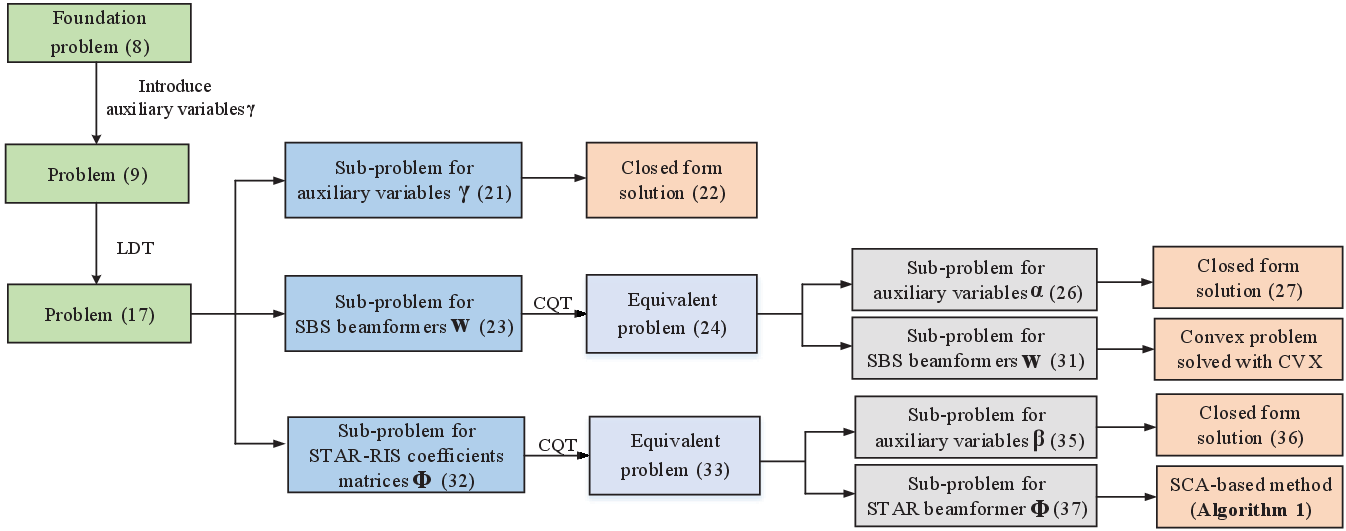}
    \caption{{The algorithm flow diagram for Algorithm 2}}
    \label{flow}
    \end{figure*}
{\textbf{Algorithm~\ref{algorithm2}} summarize the proposed BCD algorithm for problem~\eqref{problem:sum_rate_LDR}. A diagrammatic representation of \text{Algorithm~\ref{algorithm2}} is given in Fig.~\ref{flow}. {Generally, the sub-problems in the BCD algorithm are required to be solved to their optimal solutions to guarantee the convergence of the overall algorithm~\cite{doi:10.1137/120891009}. However, the STAR-RIS transmission and reflection matrices optimization problem~\eqref{problem:phi} in the BCD iteration is only solved to its stationary point. Therefore, the convergence analysis of traditional BCD algorithm cannot be applied here. The convergence of the Algorithm~\ref{algorithm2} is discussed in the following theorem. \begin{theorem}\label{proposition22}
\emph{\text{Algorithm~\ref{algorithm2}} can at least convergence to a stationary point of~\eqref{problem:sum_rate_LDR}.} 
\begin{proof}
See Appendix B.
\end{proof}
\end{theorem}}



{The complexity of the BCD Algorithm 2 can be given as follows. In Step 3, the complexity of computing the auxiliary vector $\bm{\gamma}$ is $\mathcal{O}\left(L((LM_s+2)U_s+LU_s^2+U_s^3)\right)$. In Step 4, the complexity of computing the auxiliary vector $\bm{\alpha}$ is $\mathcal{O}\left(L\left(\left(LM_s+1\right)U_s+LU_s^2+U_s^3\right)\right)$. In Step 5, the complexity of solving the convex quadratically constrained quadratic program with the interior-point method is $\mathcal{O}\left(\left(K+1\right)^{\frac{1}{2}}\left(\left(K+1\right)L^2M_s^2+ L^3M_s^3\right)\right)$~\cite{nesterov1994interior}. In Step 6, the complexity of computing the auxiliary vector $\bm{\beta}$ is $\mathcal{O}\left(L\left(U_s +LU_s^2+U_s^3\right)\right)$. In Step 7, the non-convex problem is solved with the SCA-based algorithm, which is summarized in Algorithm~\ref{algorithm1}. In each iteration of Algorithm~\ref{algorithm1}, we solve a semidefinite programming problem with the worst-case computational complexity of $\mathcal{O}\left(\text{max}\{K+2,2N+1\}^4\left(2N+1\right)^{\frac{1}{2}}\right)$~\cite{5447068}. Furthermore, since the STAR-RIS is generally equipped with numerous elements, the worst-case complexity of solving problem~\eqref{problem:phi} in Step 7 is $\mathcal{O}\left(\log\left(\frac{1}{\epsilon_{\ref{algorithm1},\text{SCA}}}\right)\left(2N+1\right)^{4.5}\right)$, where $\epsilon_{\ref{algorithm1},\text{SCA}}$ is the convergence threshold for the SCA method. Finally, the overall complexity of the BCD Algorithm 2 is given in~\eqref{C1} at the top of the next page, where $\epsilon_{\ref{algorithm2},\text{BCD}}$ is the convergence threshold for the BCD algorithm.
\begin{figure*}[!t]
\normalsize
\begin{equation}\label{C1}
\begin{aligned}
O_2=\mathcal{O}\Bigg(\log(\frac{1}{\epsilon_{\ref{algorithm2},\text{BCD}}})&\Big(3L\big((\frac{2}{3}LM_s+\frac{4}{3})U_s+LU_s^2+U_s^3\big)\\
&+\left(K+1\right)^{\frac{1}{2}}\left(\left(K+1\right)L^2M_s^2+ L^3M_s^3\right)+\log(\frac{1}{\epsilon_{\ref{algorithm1},\text{SCA}}})(2N+1)^{4.5})\Big)\Bigg).
\end{aligned}
\end{equation}
\hrulefill \vspace*{0pt}
\end{figure*}}



\begin{algorithm}[t]
\caption{The BCD algorithm for solving problem~\eqref{problem:sum_rate_LDR}.}\label{algorithm2}
\begin{algorithmic}[1]
\STATE {Given $\mathbf{w}^{(0)}$, $\mathbf{\Phi}^{(0)}$, $\bm{\gamma}^{(0)}$, and $\epsilon_{\ref{algorithm2},\text{BCD}} >0$, set $l=0$.}\\
\STATE {\bf repeat: }\\
\STATE \quad Given $\mathbf{w}^{(l)}$ and $\mathbf{\Phi}^{(l)}$, obtain $\bm{\gamma}^{(l+1)}$ with~\eqref{gamma_opt}.\\
\STATE \quad Given $\mathbf{w}^{(l)}$, $\mathbf{\Phi}^{(l)}$, and $\bm{\gamma}^{(l+1)}$, obtain $\bm{\alpha}^{(l+1)}$ with~\eqref{alpha}.\\
\STATE \quad Given $\mathbf{\Phi}^{(l)}$, $\bm{\gamma}^{(l+1)}$, and $\bm{\alpha}^{(l+1)}$, solve $\mathbf{w}^{(l+1)}$.\\
\STATE \quad Given $\mathbf{\Phi}^{(l)}$, $\mathbf{w}^{(l+1)}$, and $\bm{\gamma}^{(l+1)}$, obtain $\bm{\beta}^{(l+1)}$ with~\eqref{beta}.\\
\STATE \quad Given $\mathbf{w}^{(l+1)}$, $\bm{\gamma}^{(l+1)}$, and $\bm{\beta}^{(l+1)}$, solve $\mathbf{\Phi}^{(l+1)}$ with Algorithm 1.\\
\STATE \quad Set $l =l+1$.\\
\STATE {\bf until} the fractional increase of~\eqref{obj:sum_rate_LDR} is below~$\epsilon_{\ref{algorithm2},\text{BCD}}$.
\end{algorithmic}
\end{algorithm}

\section{Joint Beamforming Algorithm with Coupled STAR-RIS Model} 
{For STAR-RIS aided CR system where the STAR-RIS adopts the {coupled phase-shift model}, the joint beamforming design can also use the alternating optimization strategy. Moreover, the problem reformulation and the design for active beamformers are identical to the CR system with the {independent phase-shift model}. Thus, the derivations for updating $\bm{\gamma}$, $\bm{\alpha}$, $\mathbf{w}$, and $\bm{\beta}$ in each iteration are identical to those in Section III and are therefore omitted.}
Then, we focus on optimize $\mathbf{\Phi}/\mathbf{v}$ under the coupled phase shift constraints with given $\bm{\gamma}$, $\bm{\alpha}$, $\mathbf{w}$, and $\bm{\beta}$. The corresponding optimization problem can be given as follows:
\begin{subequations}\label{problem:phi_cou}
    \begin{align}
        \label{obj:phi_cou}        
        \min_{\mathbf{v}}\quad &  \sum_{i\in\{t,r\}}(\mathbf{v}_i^H\mathbf{C}_i\mathbf{v}_i - 2\Re(\mathbf{c}_i^H\mathbf{v}_i)) \\
        \label{constraint:STAR_cou1} 
        \mathrm{s.t.} \quad & \rho_n^t+\rho_n^r=1, \rho_n^t, \rho_n^r\in [0,1], \forall n,\\
        \label{constraint:STAR_cou2} 
        & \left|\theta_n^t-\theta_n^r\right|=\frac{\pi}{2} \ \text{or}\ \frac{3\pi}{2}, \forall n,\\
        & \eqref{RE_gamma1}.
    \end{align}
\end{subequations}
Problem~\eqref{problem:phi_cou} is non-convex due to the non-convex constraint~\eqref{constraint:STAR_cou2} where the transmission and reflection phase shifts of each STAR element are coupled. With given transmission phase shift $\theta_n^t$, the reflection phase shift, i.e., $\theta_n^r$, can only be selected form a  discrete set $\{\theta_n^t \pm \frac{\pi}{2}, \theta_n^t \pm \frac{3\pi}{2}\}$. {The PDD algorithm is an optimization framework designed to solve non-convex problems with variables that are nonlinearly coupled within non-convex constraints~\cite{9120361,9119203,9935266}. Therefore, it is a natural choice to adopt the PDD algorithm to handle the STAR-RIS with coupled phase shifts.}

To facilitate the following derivations, we define vectors 
\begin{equation} 
\begin{aligned}
\tilde{\mathbf{v}}_i=\left[\sqrt{\tilde{\rho}_1^i}e^{j\tilde {\theta}_1^i}, \sqrt{\tilde{\rho}_2^i}e^{j\tilde {\theta}_2^i}, \cdots, \sqrt{\tilde{\rho}_N^i}e^{j\tilde {\theta}_N^i}\right]^T, \forall i \in\{t,r\}.
\end{aligned}
\end{equation}
Problem~\eqref{problem:phi_cou} is equivalent to following problem
\begin{subequations}\label{problem:phi_cou_aux}
    \begin{align}
        \label{obj:phi_cou_aux}        
        \min_{\mathbf{v},\tilde{\mathbf{v}}}\quad &  \sum_{i\in\{t,r\}}(\mathbf{v}_i^H\mathbf{C}_i\mathbf{v}_i - 2\Re(\mathbf{c}_i^H\mathbf{v}_i)) \\
        \label{constraint:STAR_cou1_aux} 
        \mathrm{s.t.} \quad & \tilde{\mathbf{v}}_n^i={\mathbf{v}}_n^i, \forall i \in\{t,r\},\\
        \label{constraint:STAR_cou2_aux}
        & \tilde{\rho}_n^t+\tilde{\rho}_n^r=1, \tilde{\rho}_n^t, \tilde{\rho}_n^r\in [0,1], \forall n,\\
        \label{constraint:STAR_cou3_aux}       
        & \left|\tilde{\theta}_n^t-\tilde{\theta}_n^r\right|=\frac{\pi}{2} \ \text{or}\ \frac{3\pi}{2}, \forall n,\\
        & \eqref{RE_gamma1}.
    \end{align}
\end{subequations}
where $\tilde{\mathbf{v}}=\{\tilde{\mathbf{v}}_t,\tilde{\mathbf{v}}_r\}$. The PDD-based algorithm is adopted to solve this problem with a two-layer iteration, where the auxiliary variables are updated in the outer loop following the operation provided in~\cite{9120361}, and variables $ \mathbf{v}$ and $\tilde{\mathbf{v}}$ are optimized in the inner loop. 

With auxiliary variables Lagrangian dual vectors ${\bm{\tau }} = \{{\bm{\tau }}_t, {\bm{\tau }}_r\}$ and penalty factors $\eta = \{\eta_t, \eta_r\}$, the inner loop augmented Lagrangian (AL) problem for problem~\eqref{problem:phi_cou_aux} is given by 
\begin{subequations}\label{problem:PDD}
    \begin{align} 
        \label{problem:PDD_obj}       
       \min_{\mathbf{v},\tilde{\mathbf{v}}}   &  \sum_{i\in\{t,r\}}\left(\mathbf{v}_i^H\mathbf{C}_i\mathbf{v}_i - 2\Re(\mathbf{c}_i^H\mathbf{v}_i)+\frac{1}{2\eta}\|\tilde{\mathbf{v}}_i-{\mathbf{v}}_i+\eta{{\bm{\tau}}_i} \|_2^2\right) \\
        \mathrm{s.t.}   &\quad \eqref{RE_gamma1}, \eqref{constraint:STAR_cou2_aux}, \eqref{constraint:STAR_cou3_aux}.
    \end{align}
\end{subequations}
The variables in the AL problem are separable given specific dual vectors and penalty factors. Problem~\eqref{problem:PDD} is addressed using the BCD approach with two sets of variables: $\mathbf{v}$ and $\tilde{\mathbf{v}}$.
\subsection{Optimizing $\mathbf{v}$ for given $\tilde{\mathbf{v}}$} With fixed auxiliary vectors $\tilde{\mathbf{v}}$, the optimization problem for ${\mathbf{v}}$ can be given as follows:
\begin{subequations}\label{problem:PDD1}
    \begin{align}        
       \min_{\mathbf{v}}  &  \sum_{i\in\{t,r\}}\left(\mathbf{v}_i^H\mathbf{C}_i\mathbf{v}_i - 2\Re(\mathbf{c}_i^H\mathbf{v}_i)+\frac{1}{2\eta}\|\tilde{\mathbf{v}}_i-{\mathbf{v}}_i+\eta{{\bm{\tau}}_i} \|_2^2\right)\\
       \mathrm{s.t.} &\quad   \eqref{RE_gamma1}.
    \end{align}
\end{subequations}
Problem~\eqref{problem:PDD1} is a convex problem and its optimal solution can be obtained via CVX~\cite{grant2014cvx}.
\subsection{Optimizing $\tilde{\mathbf{v}}$ for given $\mathbf{v}$} With fixed ${\mathbf{v}}$, the optimization problem for $\tilde{\mathbf{v}}$ can be given as follows:
\begin{subequations}\label{problem:PDD2}
    \begin{align}        
       \min_{\tilde{\mathbf{v}}}   \quad&  \sum_{i\in\{t,r\}}\|\tilde{\mathbf{v}}_i-{\bm{\upsilon}}_i \|_2^2 \\
       \mathrm{s.t.}   \quad& \eqref{constraint:STAR_cou2_aux}, \eqref{constraint:STAR_cou3_aux},
    \end{align}
\end{subequations}
where ${\bm{\upsilon}}_i={\mathbf{v}}_i-\eta{{\bm{\tau}}_i}=[\upsilon_{1,i},\upsilon_{2,i},\cdots,\upsilon_{N,i}]^T, \forall i \in\{t,r\}$. Using the constraint $\tilde{\rho}_n^t+\tilde{\rho}_n^r=1$, the objective function of problem~\eqref{problem:PDD2} can be substituted with $-\sum_{i\in\{t,r\}}\Re({\bm{\upsilon}}_i^H\tilde{\mathbf{v}}_i)$. Thus, problem~\eqref{problem:PDD2} can be parallelly separated into $n$ optimization problem, where the problem corresponding to the $n$-th STAR element is given by
\begin{subequations}\label{problem:PDD2_n}
    \begin{align}        
       \max_{\tilde{\rho}_n^t, \tilde{\rho}_n^r, \tilde{\theta}_n^t, \tilde{\theta}_n^r}   \quad&  \sum_{i\in\{t,r\}}\sqrt{\tilde{\rho}_n^i}\Re(\upsilon_{n,i}^*e^{j\tilde{\theta}_n^i}) \\
       \label{constraint:STAR_cou1_n} 
        \mathrm{s.t.} \quad & \tilde{\rho}_n^t+\tilde{\rho}_n^r=1, \tilde{\rho}_n^t, \tilde{\rho}_n^r\in [0,1],\\
        \label{constraint:STAR_cou2_n}       
        & \left|\tilde{\theta}_n^t-\tilde{\theta}_n^r\right|=\frac{\pi}{2} \ \text{or}\ \frac{3\pi}{2}.
    \end{align}
\end{subequations}
Then problem~\eqref{problem:PDD2_n} is solved with BCD method where the auxiliary amplitude coefficients $\tilde{\rho}_n^t$ and $ \tilde{\rho}_n^r$ are optimized with fixed auxiliary phase shifts $\tilde{\theta}_n^t$ and $ \tilde{\theta}_n^r$, and vise versa.

\subsubsection{Optimizing $\tilde{\rho}_n^t$ and $ \tilde{\rho}_n^r$ for given $\tilde{\theta}_n^t$ and $ \tilde{\theta}_n^r$}With given auxiliary phase shifts $\tilde{\theta}_n^t$ and $ \tilde{\theta}_n^r$, the sub-problem with respect to auxiliary amplitude coefficients can be formulated as follows:
\begin{subequations}\label{problem:PDD2_n_am}
    \begin{align}        
       \max_{\tilde{\rho}_n^t}   \quad&  \sqrt{\tilde{\rho}_n^t}\psi_n^t+\sqrt{1-\tilde{\rho}_n^t}\psi_n^r ,
    \end{align}
\end{subequations}
where $\psi_n^i = \Re(\upsilon_{n,i}^*e^{j\tilde{\theta}_n^i}), \forall n,i.$ We notice that when the value of $\psi_n^t$ and $\psi_n^r$ are not both positive, the solution of problem~\eqref{problem:PDD2_n_am} always located at the boundary of the 
feasible region. When both the value of $\psi_n^t$ and $\psi_n^r$ are positive, problem~\eqref{problem:PDD2_n_am} can be transformed into following form:
\begin{subequations}\label{problem:PDD2_n_am1}
    \begin{align}        
       \max_{\|\tilde{\bm{\rho}}_n\|_2 \le 1}   \quad&  \bm{\psi}_n^T\tilde{\bm{\rho}}_n,
    \end{align}
\end{subequations}
where $\bm{\psi}_n = [{\psi_n^t}, {\psi_n^r}]^T$  and $\tilde{\bm{\rho}}_n= [\sqrt{\tilde{\rho}_n^t}, \sqrt{\tilde{\rho}_n^r}]^T$. The optimization is equivalent to finding the largest projection on $\bm{\psi}_n$ inside a unit circle. Thus the optimal solution to $\tilde{\bm{\rho}}_n$ is 
\begin{equation}
    \tilde{\bm{\rho}}_n = \frac{\bm{\psi}_n}{\|\bm{\psi}_n\|_2}.
\end{equation}
Then we have the following proposition:
\begin{proposition}\label{proposition1}
\emph{With given auxiliary phase shifts, optimal auxiliary amplitude coefficients can be solved in the following closed form}
\begin{subnumcases}{ ({\sqrt{\tilde{\rho}_n^t}}^{\text{opt}}, {\sqrt{\tilde{\rho}_n^r}}^{\text{opt}}) =  }
   (0,0),\  \text{if} \  {\psi_n^t},{\psi_n^r} \le 0,\\
   (1,0),\  \text{if} \  {\psi_n^t}\ge 0, {\psi_n^r}\le 0,\\
   (0,1),\  \text{if} \  {\psi_n^t}\le 0, {\psi_n^r}\ge 0,\\
   (\frac{{\psi_n^t}}{\|\bm{\psi}_n\|_2},\frac{{\psi_n^r}}{\|\bm{\psi}_n\|_2}),\  \text{if} \  {\psi_n^t},{\psi_n^r} \ge 0.
\end{subnumcases}
\end{proposition}

\subsubsection{Optimizing $\tilde{\theta}_n^t$ and $ \tilde{\theta}_n^r$ for given $\tilde{\rho}_n^t$ and $ \tilde{\rho}_n^r$} We notice that constraint~\eqref{constraint:STAR_cou2_n} is equivalent to
\begin{equation}
     e^{j\tilde{\theta}_n^t} = je^{j\tilde{\theta}_n^r}\  \text{or}\  -je^{j\tilde{\theta}_n^r}.
 \end{equation} 
With given auxiliary amplitude coefficients ${\rho}_n^t$ and $ {\rho}_n^r$, the sub-problem with respect to auxiliary phase shifts can be further split as follows:
\begin{subequations}\label{problem:PDD2_n_phj}
    \begin{align}        
       \max_{\tilde{\theta}_n^t, \tilde{\theta}_n^r}   \quad&  \sum_{i\in\{t,r\}}\sqrt{\tilde{\rho}_n^i}\Re(\upsilon_{n,i}^*e^{j\tilde{\theta}_n^i}) \\
       \label{constraint:STAR_cou1_n_j} 
        \mathrm{s.t.} \quad & e^{j\tilde{\theta}_n^t} = je^{j\tilde{\theta}_n^r}.
    \end{align}
\end{subequations}
\begin{subequations}\label{problem:PDD2_n_phjj}
    \begin{align}        
       \max_{\tilde{\theta}_n^t, \tilde{\theta}_n^r}   \quad&  \sum_{i\in\{t,r\}}\sqrt{\tilde{\rho}_n^i}\Re(\upsilon_{n,i}^*e^{j\tilde{\theta}_n^i}) \\
       \label{constraint:STAR_cou1_n_jj} 
        \mathrm{s.t.} \quad & e^{j\tilde{\theta}_n^t} = -je^{j\tilde{\theta}_n^r}.
    \end{align}
\end{subequations}
We solve both optimization problems and then substitute the solutions into the objective function of problem~\eqref{problem:PDD2_n} to find the optimal solution.
\begin{algorithm}[tbp]
\caption{PDD-based algorithm for solving problem~\eqref{problem:phi_cou_aux}}\label{algorithm3}
\begin{algorithmic}[1]
\STATE {Given ${\bm{\tau}}_i^{(0)}, \forall i$, $\eta^{(0)}$, ${p}^{(0)}$,  $\mu$, $\epsilon_{\ref{algorithm3},\text{BCD}}$, $\epsilon_{\ref{algorithm3},\text{PDD}}$ and  set $n=0$.}
\STATE {\bf repeat: outer loop}
\STATE \quad Get AL problem~\eqref{problem:PDD} with regard to ${\bm{\tau}}_i^{(n)}, \forall i$, and~${\eta}^{(n)}$.\STATE \quad {\bf repeat: inner loop}
\STATE \quad\quad Fix $\tilde{\mathbf{v}}$, update $\mathbf{v}$ by solving problem~\eqref{problem:PDD1}.
\STATE \quad\quad Fix $\mathbf{v}$ and $\tilde{\eta}_n^i$, update $\tilde{\theta}_n^i$, with \textbf{Proposition~\ref{proposition1}}.
\STATE \quad\quad Fix $\mathbf{v}$ and $\tilde{\theta}_n^i$, update $\tilde{\eta}_n^i$, with \textbf{Proposition~\ref{proposition2}}.
\STATE \quad {\bf until} the fractional increase of~\eqref{problem:PDD_obj} is below $\epsilon_{\ref{algorithm3},\text{BCD}}$.
\STATE \quad Obtain $\mathbf{v}^{(n+1)}$ and $\tilde{\mathbf{v}}^{(n+1)}$.
\STATE \quad {\bf if} $\max\big(\|\bar{\mathbf{v}}_i^{(n+1)}\|_\infty, \forall i\big) \le p^{(n)}$ {\bf then}
\STATE \quad \quad Update ${\bm{\tau}}_i$ with ${\bm{\tau}}_i^{(n+1)}={\bm{\tau}}_i^{(n)}+\frac{1}{\eta^{(n)}}(\bar{\mathbf{v}}_i^{(n+1)}), \forall i$.
\STATE \quad \quad Keep the penalty factor, $\eta^{(n+1)}=\eta^{(n)}$.\STATE \quad {\bf else}
\STATE \quad \quad Update $\eta$ with $\eta^{(n+1)}=\mu\eta^{(n)}$.
\STATE \quad \quad Lagrangian dual vectors remains, ${\bm{\tau}}_i^{(n+1)}={\bm{\tau}}_i^{(n)}, \forall i$.
\STATE \quad {\bf end if}
\STATE \quad Set $p^{(n+1)}=0.9\max\big(\|\bar{\mathbf{v}}_i^{(n+1)}\|_\infty, \forall i\big)$, $n=n+1$.
\STATE {\bf until} the constraint violation is below threshold $\epsilon_{\ref{algorithm3},\text{PDD}}$.
\end{algorithmic}
\end{algorithm}
Let $\upsilon_{n,i}=a_{n,i}+jb_{n,i}$, where $a_{n,i}=\Re({\upsilon_{n,i}})$ and $b_{n,i}=\Im({\upsilon_{n,i}})$. With equation $e^{j\tilde{\theta}_n^i}=\cos(\tilde{\theta}_n^i)+j\sin(\tilde{\theta}_n^i)$, the problem~\eqref{problem:PDD2_n_phj} can be reformulated as follows:
\begin{subequations}\
    \begin{align}        
       \max_{\tilde{\theta}_n^t}   \quad&  c_{n,1}\cos(\tilde{\theta}_n^t)+c_{n,2}\sin(\tilde{\theta}_n^t),
    \end{align}
\end{subequations}
where $c_{n,1}=\sqrt{\tilde{\rho}_n^t}a_{n,t}-\sqrt{\tilde{\rho}_n^r}b_{n,r}$ and $c_{n,2}=\sqrt{\tilde{\rho}_n^t}b_{n,t}+\sqrt{\tilde{\rho}_n^r}a_{n,r}$. The optimal solution to problem~\eqref{problem:PDD2_n_phj} can be given in closed form
\begin{equation}
    {\tilde{\theta}_n^{t-}}= \angle \Bigg(\cos\bigg(\frac{c_{n,1}}{\sqrt{c_{n,1}^2+c_{n,2}^2}}\bigg)+j\sin\bigg(\frac{c_{n,2}}{\sqrt{c_{n,1}^2+c_{n,2}^2}}\bigg)\Bigg),
\end{equation}
\begin{equation}
    {\tilde{\theta}_n^{r-}}={\tilde{\theta}_n^{t-}}-\frac{\pi}{2}.
\end{equation}
Since problem~\eqref{problem:PDD2_n_phj} and~\eqref{problem:PDD2_n_phjj} have similar structures, we omit the derivations for the solution to problem~\eqref{problem:PDD2_n_phjj} for brevity. The optimal solution to problem~\eqref{problem:PDD2_n_phjj} can be given in closed form
\begin{equation}
    {\tilde{\theta}_n^{t+}}= \angle \Bigg(\cos\bigg(\frac{\hat{c}_{n,1}}{\sqrt{\hat{c}_{n,1}^2+\hat{c}_{n,2}^2}}\bigg)+j\sin\bigg(\frac{\hat{c}_{n,2}}{\sqrt{\hat{c}_{n,1}^2+\hat{c}_{n,2}^2}}\bigg)\Bigg),
\end{equation}
\begin{equation}
    {\tilde{\theta}_n^{r+}}={\tilde{\theta}_n^{t+}}+\frac{\pi}{2},
\end{equation}
where $\hat{c}_{n,1}=\sqrt{\tilde{\rho}_n^t}a_{n,t}+\sqrt{\tilde{\rho}_n^r}b_{n,r}$ and $\hat{c}_{n,2}=\sqrt{\tilde{\rho}_n^t}b_{n,t}-\sqrt{\tilde{\rho}_n^r}a_{n,r}$. Then we have the following proposition:
\begin{proposition}\label{proposition2}
\emph{With given auxiliary amplitude coefficients, optimal auxiliary phase shifts can be solved as}
\begin{subequations}\label{problem:PDD2_n_pharg}
    \begin{align}        
       (\tilde{\theta}_n^{t,\text{opt}},\tilde{\theta}_n^{r,\text{opt}})=\argmax_{(\tilde{\theta}_n^t,\tilde{\theta}_n^r)\in\mathcal{A}}   \quad&  \sum_{i\in\{t,r\}}\sqrt{\tilde{\rho}_n^i}\Re(\upsilon_{n,i}^*e^{j\tilde{\theta}_n^i}),
    \end{align}
\end{subequations}
\emph{where} $\mathcal{A}=\{(\tilde{\theta}_n^{t-},\tilde{\theta}_n^{r-}),(\tilde{\theta}_n^{t+},\tilde{\theta}_n^{r+})\}$. \emph{Substituting two set of solutions in $\mathcal{A}$ into~\eqref{problem:PDD2_n_pharg}, the optimal solution for the sub-problem with respect to auxiliary phase shifts can be obtained.} 
\end{proposition}
\subsection{PDD-based algorithm for solving problem~\eqref{problem:phi_cou_aux}}
\begin{algorithm}[t]
\caption{The BCD algorithm for solving problem~\eqref{problem:sum_rate_LDR}.}\label{algorithm4}
\begin{algorithmic}[1]
\STATE {Given $\mathbf{w}^{(0)}$, $\mathbf{\Phi}^{(0)}$, $\bm{\gamma}^{(0)}$, and $\epsilon_{\ref{algorithm4},\text{BCD}} >0$, set $l=0$.}\\
\STATE {\bf repeat: }\\
\STATE \quad Given $\mathbf{w}^{(l)}$ and $\mathbf{\Phi}^{(l)}$, obtain $\bm{\gamma}^{(l+1)}$ with~\eqref{gamma_opt}.\\
\STATE \quad Given $\mathbf{w}^{(l)}$, $\mathbf{\Phi}^{(l)}$, and $\bm{\gamma}^{(l+1)}$, obtain $\bm{\alpha}^{(l+1)}$ with~\eqref{alpha}.\\
\STATE \quad Given $\mathbf{\Phi}^{(l)}$, $\bm{\gamma}^{(l+1)}$, and $\bm{\alpha}^{(l+1)}$, solve $\mathbf{w}^{(l+1)}$.\\
\STATE \quad Given $\mathbf{\Phi}^{(l)}$, $\mathbf{w}^{(l+1)}$, and $\bm{\gamma}^{(l+1)}$, obtain $\bm{\beta}^{(l+1)}$ with~\eqref{beta}.\\
\STATE \quad Given $\mathbf{w}^{(l+1)}$, $\bm{\gamma}^{(l+1)}$, and $\bm{\beta}^{(l+1)}$, solve $\mathbf{\Phi}^{(l+1)}$ with Algorithm 3.\\
\STATE \quad Set $l =l+1$.\\
\STATE {\bf until} the fractional increase of~\eqref{obj:sum_rate_LDR} is below $\epsilon_{\ref{algorithm4},\text{BCD}}$.
\end{algorithmic}
\end{algorithm}

{\textbf{Algorithm~\ref{algorithm3}} is a summary of the PDD-based algorithm for addressing the subproblem with respect to $\mathbf{v}$ for STAR-RIS with coupled phase shifts, where $\bar{\mathbf{v}}_i^{(n)}=\tilde{\mathbf{v}}_i^{(n)}-{\mathbf{v}}_i^{(n)}, \forall i \in \{t,r\}$. {Algorithm~\ref{algorithm3} can converge to the Karush–Kuhn–Tucker (KKT) optimal solution of the coupled phase-shift STAR-RIS coefficients optimization problem~\eqref{problem:phi_cou_aux}. We refer readers to the proof of the convergence of the PDD-based algorithm in work~\cite{9935266}.} The complexity of the Algorithm~\ref{algorithm3} can be given as follows. In Step 5 of Algorithm~\ref{algorithm3}, the complexity of solving the convex quadratically constrained quadratic program with the interior-point method is $\mathcal{O}\left(4K^{\frac{3}{2}}N^2+8K^{\frac{1}{2}}N^3\right)$. In Step 6 of Algorithm~\ref{algorithm3}, the complexity of solving auxiliary phase shifts with closed-form solutions given in {Proposition 1} is  $\mathcal{O}\left(4N\right).$ In Step 7 of Algorithm~\ref{algorithm3}, the complexity of solving auxiliary amplitude coefficients with closed-form solutions given in {Proposition 2} is $\mathcal{O}\left(2N\right).$}

\subsection{The Overall Algorithm}
{{\textbf{Algorithm~\ref{algorithm4}} summarize the proposed BCD algorithm for problem~\eqref{problem:sum_rate_LDR}. Similarly to Algorithm~\ref{algorithm4}, the coupled phase-shift STAR-RIS transmission and reflection matrices optimization problem~\eqref{problem:phi_cou} in the BCD iteration is only solved to its stationary point. Therefore, the convergence analysis of traditional BCD algorithm cannot be directly adopted here. The convergence of the Algorithm~\ref{algorithm4} is discussed in the following theorem.

\begin{theorem}\label{proposition33}
\emph{\text{Algorithm~\ref{algorithm4}} can at least obtain a stationary point of~\eqref{problem:sum_rate_LDR}.} 
\begin{proof}
See Appendix C.
\end{proof}
\end{theorem}}

The complexity of the BCD Algorithm~\ref{algorithm4} can be given as follows. In Steps 3-6 of Algorithm~\ref{algorithm4}, the complexity of computing the active beamforming vectors $\mathbf{w}$, auxiliary vectors $\bm{\gamma}$, $\bm{\alpha}$, and $\bm{\beta}$ is the same to that in Algorithm~\ref{algorithm2}. In Step 7 of Algorithm~\ref{algorithm4}, the non-convex problem is solved with the PDD-based algorithm, which is summarized in Algorithm~\ref{algorithm3}. Finally, the overall complexity of the BCD Algorithm~\ref{algorithm4} is given in~\eqref{C2} at the top of the next page,
\begin{figure*}[!t]
\normalsize
\begin{equation}\label{C2}
\begin{aligned}
O_4={{\cal O}\Bigg(\log (\frac{1}{{{\epsilon_{\ref{algorithm4},{\rm{BCD}}}}}})\Big(3L\big( {(\frac{2}{3}L{M_s} + \frac{4}{3}){U_s} + LU_s^2 + U_s^3} \big)} &{+ {\left( {K + 1} \right)^{\frac{1}{2}}}\left( {\left( {K + 1} \right){L^2}M_s^2 + {L^3}M_s^3} \right)}\\ 
&{+ \log (\frac{1}{{{\epsilon_{\ref{algorithm3},{\rm{BCD}}}}}})\log (\frac{1}{{{\epsilon_{\ref{algorithm3},{\rm{PDD}}}}}})({6N + 4{K^{\frac{3}{2}}}{N^2} + 8{K^{\frac{1}{2}}}{N^3}} )\Big)\Bigg).}
\end{aligned}
\end{equation}
\hrulefill \vspace*{0pt}
\end{figure*}
where $\epsilon_{\ref{algorithm3},\text{BCD}}$, $\epsilon_{\ref{algorithm3},\text{PDD}}$, and $\epsilon_{\ref{algorithm4},\text{BCD}}$ are the convergence threshold for the BCD iteration in Algorithm~\ref{algorithm3}, the convergence threshold for the PDD iteration in Algorithm~\ref{algorithm3}, and the convergence threshold for the BCD iteration in Algorithm~\ref{algorithm4}, respectively.

\section{Simulation Results} \label{sec:results}
To validate the effectiveness of the proposed methods, we present simulation results obtained from Monte Carlo simulations in this section.
\begin{figure}[!htbp]
\centering
\includegraphics[width=0.45\textwidth]{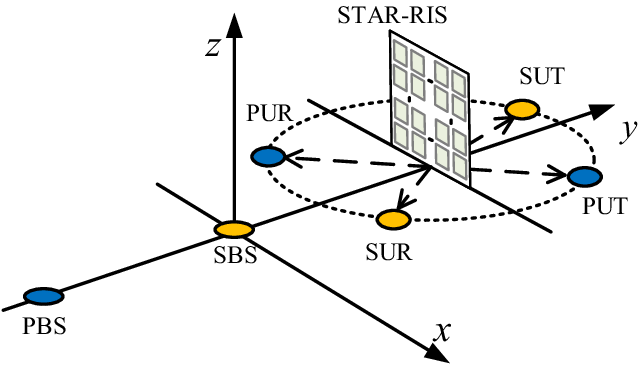}
\caption{Simulation setup for the STAR-RIS aided CR network.}
\label{setup}
\end{figure}

\subsection{Simulation Setup}
The considered STAR-RIS aided CR system is shown in Fig.~\ref{setup}, with the PBS, the SBS, and the STAR-RIS placed at $(0,-50)$, $(0,0)$, and $(0,30)$ in meter (m). All PUs and SUs lie on the circle around the STAR-RIS with a radius of $10$~m with random angles. It is assumed that $2$ SUs and $4$ PUs are equally separated into the two sides of the STAR-RIS. All involved channels are modeled as follows:
\begin{equation}
\mathbf{H}=\sqrt{\frac{\text{PL}\kappa}{\kappa+1}}\mathbf{H}_\text{LoS}+\sqrt{\frac{\text{PL}}{\kappa+1}}\mathbf{H}_\text{NLoS},
\end{equation}
where $\kappa$ is the Rician factor, $\mathbf{H}_\text{LoS}$ and $\mathbf{H}_\text{NLoS}$ represent the line-of-sight (LoS) and non-LoS (NLoS) part of the channel, respectively. $\text{PL}=C_0d^{-\alpha}$ is the path loss factor, wherein $C_0=-30dB$ is the pass loss at the reference distance $1$ m, $d$ is the propagation distance, and $\alpha$ is the path loss exponent. We assume Rician factors for all STAR-RIS related channels are set as $\kappa_r=5$, while Rician factors for other channels are set as $\kappa_d=0$. The path loss exponent from the SBS to the STAR-RIS, from the PBS to the STAR-RIS, from the STAR-RIS to the users, from the SBS to users, and from the PBS to users are set as $\alpha_{S,R}=2.1$, $\alpha_{P,R}=3.6$, $\alpha_{R,U}=2.3$, $\alpha_{S,U}=3.9$, and $\alpha_{P,U}=3.9$~\cite{9183907}. Unless specific otherwise, the PBS, SBS, PUs, and SUs are equipped with $M_p=4$, $M_s=4$, $U_p=2$, $U_s=2$ antennas, respectively. The power budget at the SBS is $20$dBm. The interference from PBS at the SUs is modeled as additive white Gaussian noise and the equivalent noise at SUs is given as $\sigma_i^2=\hat{\sigma}_i^2+\sigma_s^2$. Specifically, the thermal noise power is set as $\sigma_s^2=-100$dBm and the interference noise power can be given as:
\begin{equation}
    \hat{\sigma}_i^2=P_p(\text{LP}_{S,R}^{-1}+\text{LP}_{R,i}^{-1}), \forall i \in \{t,r\},
\end{equation}
where $P_p=30$dBm is the transmit power at the PBS. $\text{LP}_{S,R}$ and $\text{LP}_{R,i}, \forall i \in \{t,r\},$ are the path loss from the SBS to the STAR-RIS and from the STAR-RIS to the SUs, respectively. {The simulation parameters adopted in simulation are listed in Table~\ref{Simulation_Parameters}.}
 \begin{table}[h]
    \caption{{Simulation Parameters}}\label{Simulation_Parameters}
    \centering
    {\begin{tabular}{|c||c||c|}
        \hline
        $M_p$ & The number of PBS antennas & $4$\\
        \hline
        $M_s$ & The number of SBS antennas & $4$\\
        \hline
        $U_p$ & The number of PU antennas & $2$\\
        \hline
        $U_s$ & The number of SU antennas & $2$\\
        \hline
        $\kappa_r$ & Rician factor for RIS channels & $5$\\
        \hline
        $\kappa_d$ & Rician factor direct channels & $0$\\
        \hline
        $P_p$ & Max. transmit power of PBS & $30$ dBm\\
        \hline
        $P_s$ & Max. transmit power of SBS & $20$ dBm\\
        \hline
        $\alpha_{S,R}$ & SBS to RIS path loss exponent & $2.1$\\
        \hline
        $\alpha_{P,R}$ & PBS to RIS path loss exponent & $3.6$\\
        \hline
        $\alpha_{S,U}$ & SBS to user path loss exponent & $3.9$\\
        \hline
        $\alpha_{P,U}$ & PBS to user path loss exponent & $3.9$\\
        \hline
        $\alpha_{R,U}$ & STAR to user path loss exponent & $2.3$\\
        \hline  
        $\sigma_s^2$ & SU noise power & $-100$ dBm\\
        \hline  
        $C_0$ & Path loss at reference distance & $30$ dB\\
        \hline 
        \end{tabular}}
    \end{table}
\subsection{Baseline Schemes}
To verify the effectiveness and benefits of the proposed STAR-RIS aided CR system, we consider the following three baseline schemes.
\begin{enumerate}
\item {\textbf{Traditional CR Scheme:} In this baseline scheme, the system is a conventional MIMO CR system without RIS. With the provided values of $\mathbf{\Phi}_t=\mathbf{\Phi}_r=\mathbf{0}$, \text{Algorithm~\ref{algorithm1}} can address the optimization problem arising from this scheme.}

\item {\textbf{Conventional RIS Scheme:} In this baseline scheme, we adopt two adjacent conventional RISs to realize full space coverage. To provide a fair comparison, each RIS contains $N/2$ elements. Configuring $\mathbf{\rho}_n^t=\mathbf{\rho}_n^r =1, \forall n\le N/2$, and $\mathbf{\rho}_n^t=\mathbf{\rho}_n^r =0, \forall n > N/2$, algorithm form work~\cite{9599553} can be used to tackle the resulted problem.}

\item \textbf{Equal Splitting STAR-RIS Scheme:} In this baseline scheme, all STAR-RIS elements equally split the signal energy into the transmission and reflection sides. With the given $\mathbf{\rho}_n^t=\mathbf{\rho}_n^r =1/2, \forall n$, \text{Algorithm~\ref{algorithm2}} and \text{Algorithm~\ref{algorithm4}} can address the optimization problem that arises from this scheme.

\end{enumerate}

In simulation figures, legends ``STAR-RIS, Independent'' and ``STAR-RIS, Coupled'' represent the CR system with STAR-RIS under {independent phase-shift model} and {coupled phase-shift model}, respectively. ``E-STAR-RIS, Independent'' represents CR system with {equal splitting} STAR-RIS under {independent phase-shift model}. Besides, {legends ``C-RIS'' and ``T-CR'' represent the CR system with conventional RIS and no RIS, respectively.}

\subsection{Convergence Behavior of Proposed Algorithms.}
\begin{figure}[!htbp]
\centering
\includegraphics[width=0.5\textwidth]{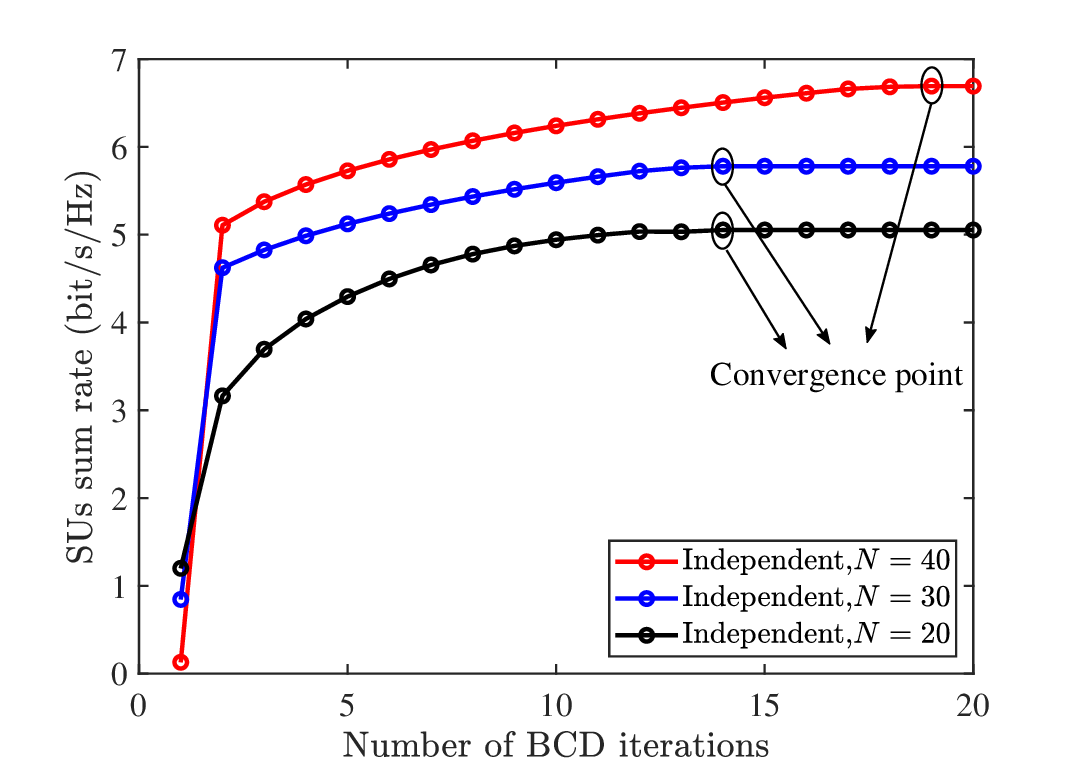}
\caption{Convergence Behavior of Proposed \textbf{Algorithm~\ref{algorithm2}}.}
\label{convergence_in}
\end{figure}
\begin{figure}[!htbp]
\centering
\includegraphics[width=0.5\textwidth]{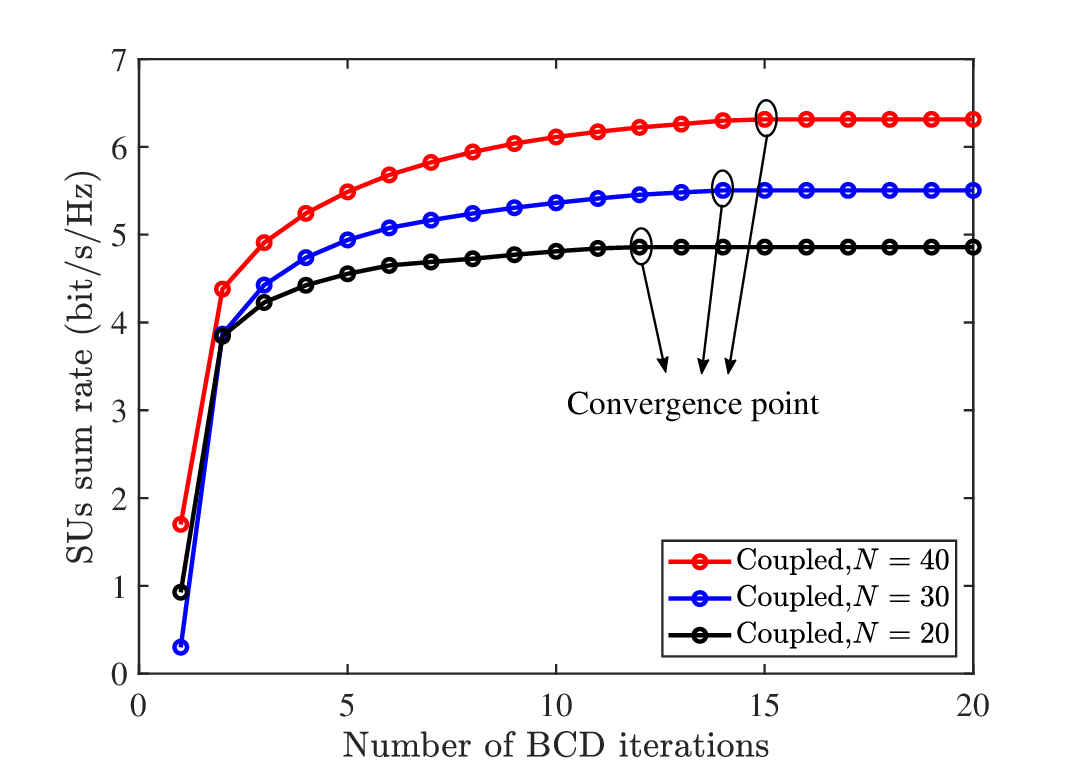}
\caption{The sum rate versus the number of BCD iterations of \text{Algorithm~\ref{algorithm4}}.}
\label{convergence_co}
\end{figure}
\begin{figure}[!htbp]
\centering
\includegraphics[width=0.5\textwidth]{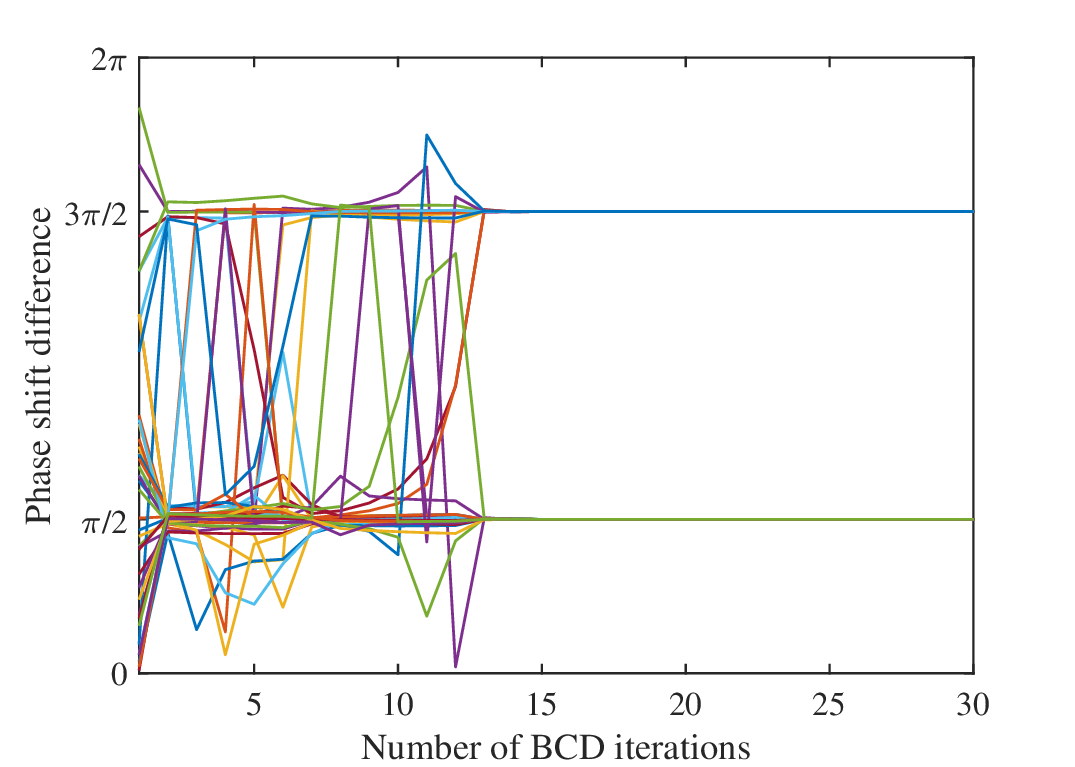}
\caption{The phase differences versus the number of BCD iterations of \text{Algorithm~\ref{algorithm4}}.}
\label{convergence_co_phase}
\end{figure}

The convergence behaviour of suggested \text{Algorithm~\ref{algorithm2}} and \text{Algorithm~\ref{algorithm4}} is examined in Fig.~\ref{convergence_in}-\ref{convergence_co_phase}. The power budget at the SBS is set as $20$~dBm and the IT threshold for PUs is set as $-90$~dBm. Fig.~\ref{convergence_in} depicts the SUs sum rate versus the number of BCD iterations under {independent phase-shift model} for three different STAR-RIS element numbers, i.e., $N=20, 30, 40$. For all STAR-RIS element numbers, the proposed algorithms converge within $20 $ iterations. The SUs sum rate under the {coupled phase-shift model} is shown against the number of BCD iterations in Fig.~\ref{convergence_co}. It is shown that the sum rate converges to a stationary point for all STAR element numbers within $15$ iterations. What’s more, Fig.~\ref{convergence_co_phase} shows the absolute phase difference for all STAR elements versus the number of BCD iterations when the number of STAR-RIS elements is $N=40$. It can be observed that the absolute phase difference for all STAR elements converges to $\frac{\pi}{2}$ or $\frac{3\pi}{2}$ with the proposed PDD-based algorithm.

\subsection{Sum Rate Versus the Power Budget of the SBS, the Number of STAR Elements, the IT Threshold, and the Number of SUs.}
\begin{figure}[!tbp]
\centering
\includegraphics[width=0.5\textwidth]{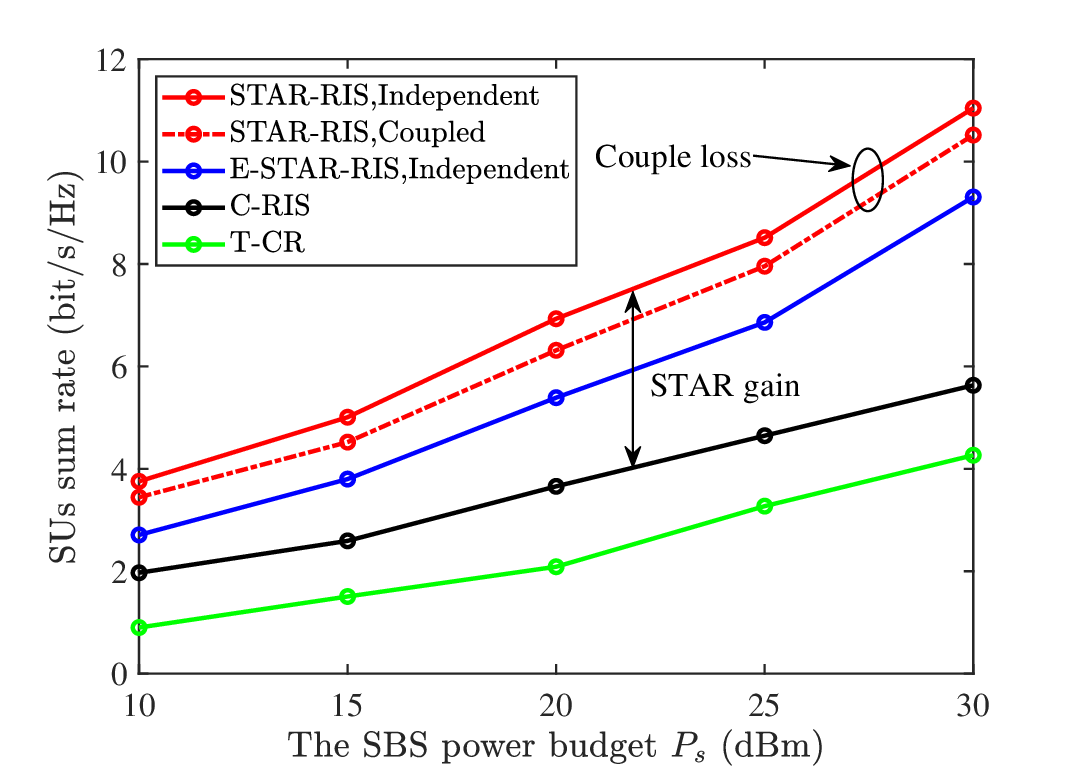}
\caption{The sum rate versus the power budget of the SBS $P_s$.}
\label{power}
\end{figure}

The sum rate of SUs versus the power budget of the SBS for each of the considered schemes is illustrated in Fig.~\ref{power}. The number of STAR elements is set to $N=40$, and the IT threshold for PUs is set at $-90$ dBm. The STAR-RIS aided schemes outperform the other schemes, attributed to the increased degrees of freedom for beamforming design provided by STAR-RIS compared to traditional RIS. Moreover, Fig.~\ref{power} indicates that, although the scheme with independent STAR-RIS outperforms the one with coupled STAR-RIS, the impact of coupled phase shifts is only a limited degradation in performance.

\begin{figure}[!tbp]
\centering
\includegraphics[width=0.5\textwidth]{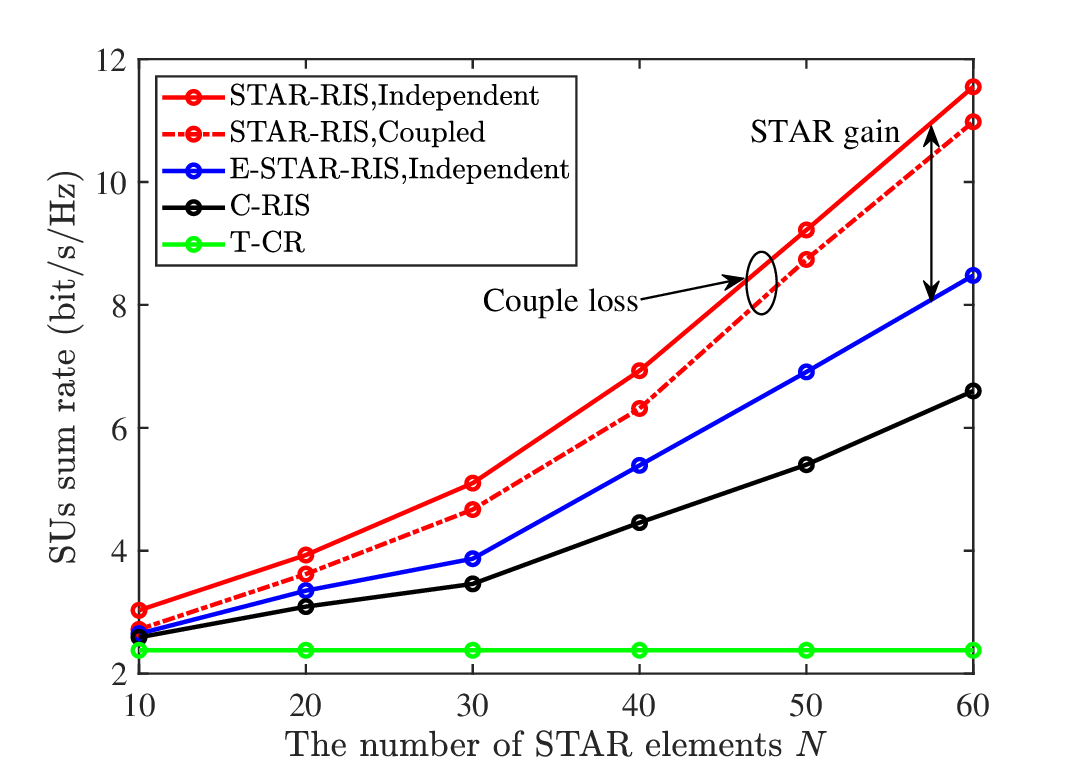}
\caption{The sum rate versus the number of STAR elements $N$.}
\label{elements}
\end{figure}
The SUs sum rate against the number of RIS elements for all schemes is displayed in Fig.~\ref{elements}. The power budget of the SBS is set at $P_s=20$ dBm, and the IT threshold for PUs is set at $-90$ dBm. The sum rate for all RIS-related schemes rises as the number of RIS elements grows, as shown in Fig.~\ref{elements}. This finding is expected as a larger passive beamforming gain can be provided with a bigger $N$. In Fig.~\ref{elements}, the STAR-RIS assisted scheme also performs better than the traditional RIS scheme, which is consistent with the analysis for Fig.~\ref{power}.

\begin{figure}[!tbp]
\centering
\includegraphics[width=0.5\textwidth]{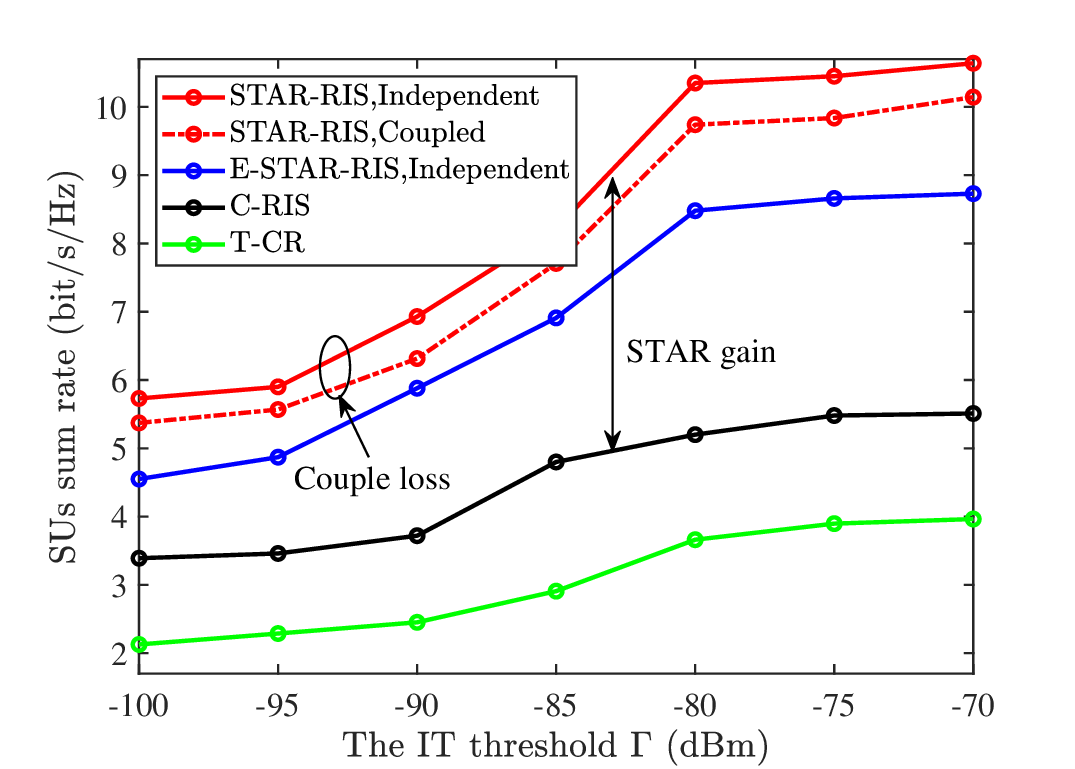}
\caption{The sum rate versus the IT threshold $\Gamma$.}
\label{fig:gamma}
\end{figure}
Fig.~\ref{fig:gamma} shows the SUs sum rate versus the IT threshold for all considered schemes. The number of STAR elements is set as $N=40$ and the power budget of the SBS is set as $P_s=20$ dBm. The CR system can achieve better performance as the IT threshold becomes higher. This is expected since the relatively higher IT threshold means the secondary system is confined more loosely, which allows the SBS to achieve better system performance. It is also noticed that when the IT constraints are strict (IT threshold under $-80$ dBm in Fig.~\ref{fig:gamma}, the sum rate of SUs is mainly confined by the IT constraint. In this region, increasing the IT threshold can effectively enhance the sum rate of SUs. On the contrary, when the IT threshold is sufficiently high (IT threshold above $-80$ dBm in Fig.~\ref{fig:gamma}, the IT constraints can be easily satisfied. In this scenario, the sum rate of SUs is primarily constrained by the power budget of the SBS. In this region, increasing the IT threshold does not effectively enhance the sum rate of SUs.

\begin{figure}[!tbp]
\centering
\includegraphics[width=0.5\textwidth]{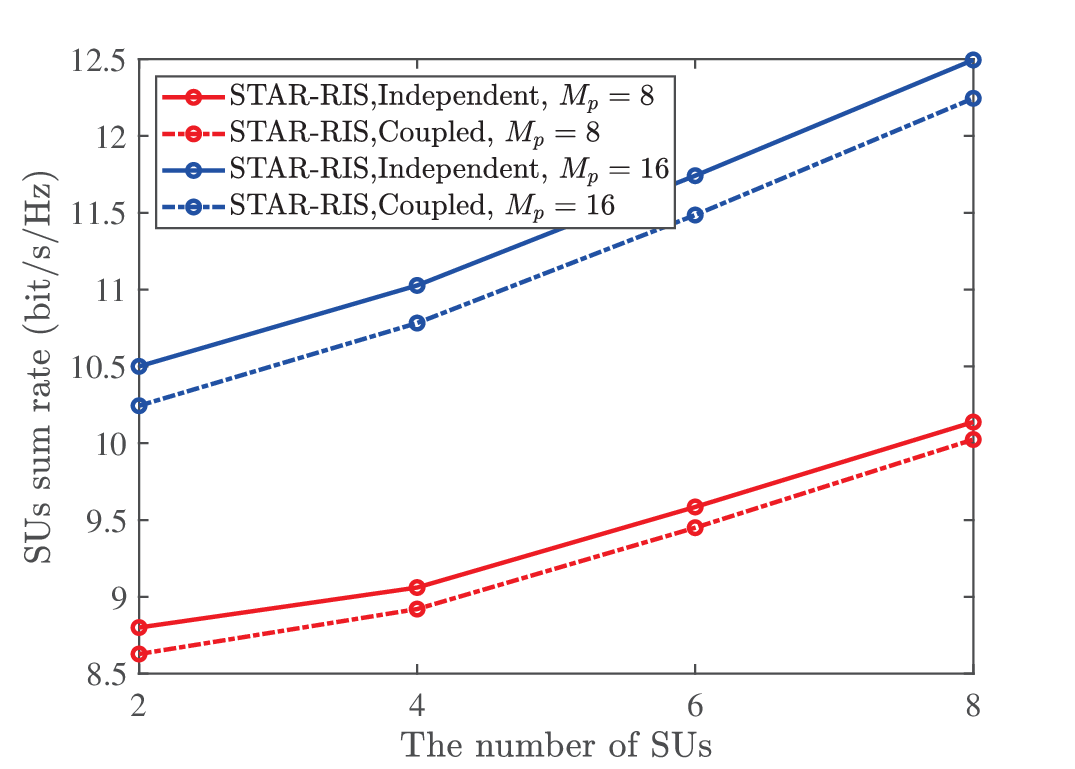}
\caption{{The sum rate versus the number of SUs.}}
\label{fig:SUs}
\end{figure}
{Fig.~\ref{fig:SUs} depicts the sum rate of the secondary users (SUs) versus the number of SUs. The number of STAR elements is set to $N=40$, and the power budget of the SBS is fixed at $P_s=20$ dBm. It is observed that as the number of SUs, $L$, increases, the sum rate of the secondary network also increases. This result is expected, as the beamforming design can exploit multiuser diversity. Furthermore, it is noted that the secondary network achieves a higher sum rate with more antennas at the SBS. This improvement can be attributed to the increased antenna diversity at the SBS, which helps combat fading and enhances the reliability of communication links, thereby leading to a higher sum rate. Additionally, it is observed that the performance improvement resulting from an increase in the number of SUs becomes more evident when the SBS is equipped with more antennas. This is because the effect of beamforming is more pronounced when the SBS has a larger antenna array.}
\section{Conclusion}
In this work, we investigated a STAR-RIS aided MIMO CR system. A pair of efficient secondary system sum rate maximization algorithms were conceived for both {independent phase-shift model} and {coupled phase-shift model}. Simulation results demonstrated that: 1) the superiority of the STAR-RIS aided CR communication system over the considered benchmark schemes; 2) the introduction of coupled phase shifts constraint in STAR-RIS resulted in limited performance degradation when compared to independent phase STAR-RIS in CR systems. In our future research, it would be intriguing to explore how STAR-RIS can achieve beamfocusing specifically over users in its near-field region, allowing for more precise cognitive radio interference management. 
{Additionally, MS STAR-RISs are also promising in STAR-RIS aided CR systems. They offer a lower manufacturing cost compared to ES STAR-RISs while still providing simultaneous full-space coverage. }

{\numberwithin{equation}{section}
\section*{Appendix~A: Proof of Theorem~\ref{proposition11}} \label{Appendix:A}
\renewcommand{\theequation}{A.\arabic{equation}}
\setcounter{equation}{0}
Define $f_{\text{a}}(\mathbf{v})$ and $f_{\text{a}}^{\text{ub}}(\mathbf{v})$ as the value of objective function of problem~\eqref{problem:phi1} and~\eqref{problem:phi_SCA} based on $\mathbf{v}$, respectively. During the $n$-th SCA iteration of \text{Algorithm~\ref{algorithm1}}, we have
\begin{equation}\label{a}
    \begin{aligned}
        f_{\text{a}}\big(\mathbf{v}^{(n)}\big) \buildrel (a) \over= f_{\text{a}}^\text{ub}\big(\mathbf{v}^{(n)}\big)
        \buildrel (b) \over\ge f_{\text{a}}^\text{ub}\big(\mathbf{v}^{(n+1)}\big) \buildrel (c) \over\ge f_{\text{a}}\big(\mathbf{v}^{(n+1)}\big),
    \end{aligned}
\end{equation}
where equality~$(a)$ is valid because the lower bound in~\eqref{v_SCA} is tight at the given point~$\mathbf{v}^{(n)}$. Thus, the approximated problem~\eqref{problem:phi_SCA} can achieve the same optimal objective function as that of problem~\eqref{problem:phi1} at the given point~$\mathbf{v}^{(n)}$. Inequality~$(b)$ is valid since the problem~\eqref{problem:phi_SCA} is solved to its optimal solution~$\mathbf{v}^{(n+1)}$ in the step 3 of Algorithm~\ref{algorithm1}. Inequality~$(c)$ is valid since the optimal objective function
of the approximation problem~\eqref{problem:phi_SCA} serves as a upper bound of that of the original problem~\eqref{problem:phi1}.

It is shown in~\eqref{a}  that in each iteration of the SCA algorithm, the objective function value of~\eqref{problem:phi1} is non-increasing. Besides, the objective function value of~\eqref{problem:phi1} is bounded below since its objective function is in the positive definite quadratic form. As a result, \text{Algorithm~\ref{algorithm1}} can converge to at least a stationary point of problem~\eqref{problem:phi1}. Since the problem~\eqref{problem:phi} and problem~\eqref{problem:phi1} are equivalent, the obtained stationary point of problem~\eqref{problem:phi1} also serves as that of problem~\eqref{problem:phi}.

\section*{Appendix~B: Proof of Theorem~\ref{proposition22}} \label{Appendix:B}
\renewcommand{\theequation}{B.\arabic{equation}}
\setcounter{equation}{0}

During the $l$-th BCD iteration of \text{Algorithm~\ref{algorithm2}}, the initial values $\left\{\bm{\gamma},\mathbf{w}, \mathbf{\Phi}\right\}$ are set to $\left\{\bm{\gamma}^{(l)},\mathbf{w}^{(l)}, \mathbf{\Phi}^{(l)}\right\}$. Since the optimal solution of $\bm{\gamma}$ is obtained with given $\mathbf{w}^{(l)}$ and $\mathbf{\Phi}^{(l)}$, the inequality
\begin{equation}\label{b1}
	f_{\text{LDT}}(\bm{\gamma}^{(l)}, \mathbf{w}^{(l)},\mathbf{\Phi}^{(l)}) \le f_{\text{LDT}}(\bm{\gamma}^{(l+1)}, \mathbf{w}^{(l)},\mathbf{\Phi}^{(l)})
\end{equation}
can be established, where $f_{\text{LDT}}(\bm{\gamma}, \mathbf{w},\mathbf{\Phi})$ is define prior to problem~\eqref{problem:sum_rate_LDR}. Similarly, the optimal solution of $\mathbf{w}$ is obtained with given $\bm{\gamma}^{(l+1)}$ and $\mathbf{\Phi}^{(l)}$, then we have
\begin{equation}\label{b2}
	f_{\text{LDT}}(\bm{\gamma}^{(l+1)}, \mathbf{w}^{(l)},\mathbf{\Phi}^{(l)}) \le f_{\text{LDT}}(\bm{\gamma}^{(l+1)}, \mathbf{w}^{(l+1)},\mathbf{\Phi}^{(l)}).
\end{equation}

In the following, the $\mathbf{v}^{(l+1)}$ (also $\mathbf{\Phi}^{(l+1)}$) is solved using Algorithm~\ref{algorithm1} with given $\bm{\gamma}^{(l+1)}$ and $\mathbf{w}^{(l+1)}$. Recall that inequality~\eqref{a} indicates that the Algorithm~\ref{algorithm1} generate non-increasing objective function value of problem~\eqref{problem:phi}. Since the Algorithm~\ref{algorithm1} is initialed with $\mathbf{v}^{(l)}$ (also $\mathbf{\Phi}^{(l)}$), we have
\begin{equation}
	f_{\text{a}}\big(\mathbf{v}^{(l)}\big)  \ge f_{\text{a}}\big(\mathbf{v}^{(l+1)}\big).
\end{equation}
Note that with fixed $\bm{\gamma}$ and $\mathbf{w}$, the solution of $\mathbf{v}$ that reduces the objective function value of problem~\eqref{problem:phi} will lead to an increase in the objective function value of problem~\eqref{problem:sum_rate_LDR}. We have
\begin{equation}\label{b3}
	f_{\text{LDT}}(\bm{\gamma}^{(l+1)}, \mathbf{w}^{(l+1)},\mathbf{\Phi}^{(l)}) \le f_{\text{LDT}}(\bm{\gamma}^{(l+1)}, \mathbf{w}^{(l+1)},\mathbf{\Phi}^{(l+1)}).
\end{equation}
Combining~\eqref{b1},~\eqref{b2} and~\eqref{b3}, we have
\begin{equation}
    \begin{aligned}
        f_{\text{LDT}}(\bm{\gamma}^{(l)}, \mathbf{w}^{(l)},\mathbf{\Phi}^{(l)}) \le f_{\text{LDT}}(\bm{\gamma}^{(l+1)}, \mathbf{w}^{(l+1)},\mathbf{\Phi}^{(l+1)}).
    \end{aligned}
\end{equation}

It is shown that in each iteration of the BCD algorithm, the objective function value of~\eqref{problem:sum_rate_LDR} is non-decreasing. Besides, the objective function value of~\eqref{problem:sum_rate_LDR} is bounded above since the power of the SBS is limited. As a result, \text{Algorithm~\ref{algorithm2}} can converge to at least a stationary point of problem~\eqref{problem:sum_rate_LDR}.
\section*{Appendix~C: Proof of Theorem~\ref{proposition33}} \label{Appendix:C}
\renewcommand{\theequation}{C.\arabic{equation}}
\setcounter{equation}{0}

During the $l$-th BCD iteration of \text{Algorithm~\ref{algorithm4}}, the initial values $\left\{\bm{\gamma},\mathbf{w}, \mathbf{\Phi}\right\}$ are set to $\left\{\bm{\gamma}^{(l)},\mathbf{w}^{(l)}, \mathbf{\Phi}^{(l)}\right\}$. With analysis similar to that of Appendix~B, the inequalities~\eqref{b1} and~\eqref{b2} are valid for \text{Algorithm~\ref{algorithm4}}. In the following, we use proof by contradiction to prove the inequality~\eqref{b3} is also valid for \text{Algorithm~\ref{algorithm4}}.

    \noindent{\textit{Proof:}}  In the $l$-th BCD iteration of \text{Algorithm~\ref{algorithm4}}, the $\mathbf{v}^{(l+1)}$ (also $\mathbf{\Phi}^{(l+1)}$) is solved using Algorithm~\ref{algorithm3} with given $\bm{\gamma}^{(l+1)}$ and $\mathbf{w}^{(l+1)}$. Assume that 
    \begin{equation}\label{c1}
        f_{\text{LDT}}(\bm{\gamma}^{(l+1)}, \mathbf{w}^{(l+1)}, \mathbf{\Phi}^{(l)}) > f_{\text{LDT}}(\bm{\gamma}^{(l+1)}, \mathbf{w}^{(l+1)}, \mathbf{\Phi}^{(l+1)})
    \end{equation} 
    holds.

    Comparing the objective function of problem~\eqref{problem:phi_cou_aux} with $\mathbf{\Phi}^{(l)}$ and $\mathbf{\Phi}^{(l+1)}$ based on~\eqref{c1}, we have 
    \begin{equation}\label{c2}
    \begin{aligned}
        \sum_{i\in\{t,r\}}&\left((\mathbf{v}_i^{(l)})^H\mathbf{C}_i\mathbf{v}_i^{(l)} - 2\Re(\mathbf{c}_i^H\mathbf{v}_i^{(l)})\right) <\\ &\sum_{i\in\{t,r\}}\left((\mathbf{v}_i^{(l+1)})^H\mathbf{C}_i\mathbf{v}_i^{(l+1)} - 2\Re(\mathbf{c}_i^H\mathbf{v}_i^{(l+1)})\right).
        \end{aligned}
    \end{equation} 
    Recall that the penalty term in the objective function of problem~\eqref{problem:PDD1} approaches $0$ with $\eta \rightarrow 0$ and equality constraint $\tilde{\mathbf{v}}_n^i={\mathbf{v}}_n^i, \forall i \in\{t,r\}$ satisfied. With this prerequisite,~\eqref{c2} indicates $\mathbf{\Phi}^{(l)}$ yields a better solution to problem~\eqref{problem:PDD1} than $\mathbf{\Phi}^{(l+1)}$. This is contradict with the fact that $\mathbf{\Phi}^{(l+1)}$ is the optimal solution to convex problem~\eqref{problem:PDD1}. Thus, the assumption is not valid and the inequality~\eqref{b3} is also valid for \text{Algorithm~\ref{algorithm4}}.

  Since the inequalities~\eqref{b1},~\eqref{b2} and~\eqref{b3} are valid for \text{Algorithm~\ref{algorithm4}}, with similar analysis to \text{Algorithm~\ref{algorithm2}}, the convergence of \text{Algorithm~\ref{algorithm4}} is guaranteed.

}
\bibliography{myref}
\bibliographystyle{IEEEtran}

\begin{IEEEbiography}[{\includegraphics[width=1in, height=1.25in,clip, keepaspectratio]{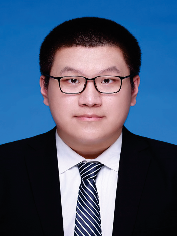}}]{Haochen Li} (Graduate Student Member, IEEE) received the B.S. degree in communication engineering from Nanjing University of Science and Technology, Nanjing, China, in 2019. He is currently pursuing the Ph.D. degree with the National Mobile Communications Research Laboratory, Southeast University, Nanjing, China. He is also a Research Associate with the School of Electronic Engineering and Computer Science, Queen Mary University of London from September 2022 to September 2024. His current research interests include massive MIMO communication, reconfigurable intelligent surface, and integrated sensing and communications.
\end{IEEEbiography}

\begin{IEEEbiography}[{\includegraphics[width=1in, height=1.25in,clip, keepaspectratio]{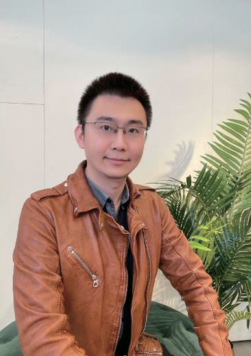}}]{Yuanwei Liu} (S'13-M'16-SM'19-F’24, \url{https://www.eee.hku.hk/~yuanwei/}) is a tenured full Professor in Department of Electrical and Electronic Engineering (EEE) at The University of Hong Kong (HKU) and a visiting professor at Queen Mary University of London (QMUL). Prior to that, he was a Senior Lecturer (Associate Professor) (2021-2024) and a Lecturer (Assistant Professor) (2017- 2021) at QMUL, London, U.K, and a Postdoctoral Research Fellow (2016-2017) at King's College London (KCL), London, U.K. He received the Ph.D. degree from QMUL in 2016.  His research interests include non-orthogonal multiple access, reconfigurable intelligent surface, near field communications, integrated sensing and communications, and machine learning. 

Yuanwei Liu is a Fellow of the IEEE, a Fellow of AAIA, a Web of Science Highly Cited Researcher, an IEEE Communication Society Distinguished Lecturer, an IEEE Vehicular Technology Society Distinguished Lecturer, the rapporteur of ETSI Industry Specification Group on Reconfigurable Intelligent Surfaces on work item of “Multi-functional Reconfigurable Intelligent Surfaces (RIS): Modelling, Optimisation, and Operation”, and the UK representative for the URSI Commission C on “Radio communication Systems and Signal Processing”. He was listed as one of 35 Innovators Under 35 China in 2022 by MIT Technology Review. He received IEEE ComSoc Outstanding Young Researcher Award for EMEA in 2020. He received the 2020 IEEE Signal Processing and Computing for Communications (SPCC) Technical Committee Early Achievement Award, IEEE Communication Theory Technical Committee (CTTC) 2021 Early Achievement Award. He received IEEE ComSoc Outstanding Nominee for Best Young Professionals Award in 2021. He is the co-recipient of the 2024 IEEE Communications Society Heinrich Hertz Award, the Best Student Paper Award in IEEE VTC2022-Fall, the Best Paper Award in ISWCS 2022, the 2022 IEEE SPCC-TC Best Paper Award, the 2023 IEEE ICCT Best Paper Award, and the 2023 IEEE ISAP Best Emerging Technologies Paper Award. He serves as the Co-Editor-in-Chief of IEEE ComSoc TC Newsletter, an Area Editor of IEEE Transactions on Communications and IEEE Communications Letters, an Editor of IEEE Communications Surveys \& Tutorials, IEEE Transactions on Wireless Communications, IEEE Transactions on Vehicular Technology, IEEE Transactions on Network Science and Engineering, and IEEE Transactions on Cognitive Communications and Networking. He serves as the (leading) Guest Editor for Proceedings of the IEEE on Next Generation Multiple Access, IEEE JSAC on Next Generation Multiple Access, IEEE JSTSP on Intelligent Signal Processing and Learning for Next Generation Multiple Access, and IEEE Network on Next Generation Multiple Access for 6G. He serves as the Publicity Co-Chair for IEEE VTC 2019-Fall, the Panel Co-Chair for IEEE WCNC 2024, Symposium Co-Chair for several flagship conferences such as IEEE GLOBECOM, ICC and VTC. He serves the academic Chair for the Next Generation Multiple Access Emerging Technology Initiative, vice chair of SPCC and Technical Committee on Cognitive Networks (TCCN).
\end{IEEEbiography}

\begin{IEEEbiography}[{\includegraphics[width=1in, height=1.25in,clip, keepaspectratio]{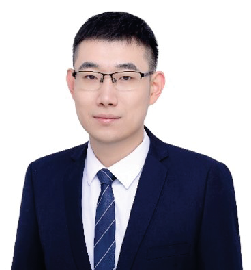}}]{Xidong Mu} (Member, IEEE, \url{https://xidongmu.github.io/}) received the Ph.D. degree in Information and Communication Engineering from the Beijing University of Posts and Telecommunications, Beijing, China, in 2022. 

He was with the School of Electronic Engineering and Computer Science, Queen Mary University of London, from 2022 to 2024, where he was a Postdoctoral Researcher. He has been a lecturer (an assistant professor) with the Centre for Wireless Innovation (CWI), School of Electronics, Electrical Engineering and Computer Science, Queen’s University Belfast, U.K. since August 2024. His research interests include non-orthogonal multiple access (NOMA), IRSs/RISs aided communications, integrated sensing and communications, semantic communications, and optimization theory. He received the IEEE ComSoc Outstanding Young Researcher Award for EMEA region in 2023, the Exemplary Reviewer Certificate of the \textsc{IEEE Transactions on Communications} in 2020 and 2022, and the \textsc{IEEE Communication Letters} in 2021-2023. He is the recipient of the 2024 IEEE Communications Society Heinrich Hertz Award, the Best Paper Award in ISWCS 2022, the 2022 IEEE SPCC-TC Best Paper Award, and the Best Student Paper Award in IEEE VTC2022-Fall. He serves as the Public Engagement and Social Networks Coordinator of IEEE ComSoC SPCC Technical Committee, the secretary of the IEEE ComSoc Next Generation Multiple Access (NGMA) Emerging Technology Initiative and the Special Interest Group (SIG) in SPCC Technical Committee on Signal Processing Techniques for NGMA. He also serves as an Editor of \textsc{IEEE Transactions on Communications}, a Guest Editor for \textsc{IEEE Transactions on Cognitive Communications and Networking} Special Issue on “Machine Learning and Intelligent Signal Processing for Near-Field Technologies”, \textsc{IEEE Internet of Things Journal} Special Issue on “Next Generation Multiple Access for Internet-of-Things”, the NGMA workshop co-chairs of IEEE WCNC 2023, IEEE PIMRC 2023, IEEE GLOBECOM 2023, and ICC 2025, and the “Mobile and Wireless Networks” symposium co-chair of IEEE GLOBECOM 2025. 
\end{IEEEbiography}

\begin{IEEEbiography}[{\includegraphics[width=1in, height=1.25in,clip, keepaspectratio]{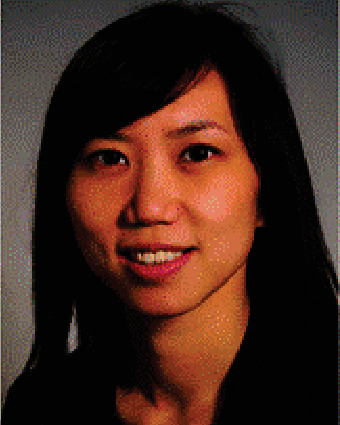}}] {Yue Chen} (Senior Member, IEEE) received the B.Eng. and M.Eng. degrees from the Beijing University of Posts and Telecommunications (BUPT), Beijing, China, in 1997 and 2000, respectively, and the Ph.D. degree from QMUL, London, U.K., in 2003.

She is currently a Professor of Telecommunications Engineering at the School of Electronic Engineering and Computer Science, Queen Mary University of London (QMUL), U.K. Her current research interests include wireless communication networks, mobile edge computing, smart energy systems, and the Internet of Things. 
\end{IEEEbiography}

\begin{IEEEbiography}[{\includegraphics[width=1in, height=1.25in,clip, keepaspectratio]{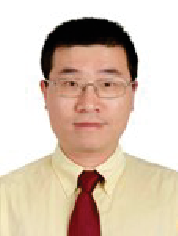}}]{Pan Zhiwen} (Member, IEEE) is currently a Professor with the National Mobile Communications Research Laboratory, Southeast University, Nanjing, China. During 2000–2001, he was involved in the research and standardization of 3G, and since 2002, he has been involved in the investigations on key technologies for IMT-A, 5G and 6G. He has authored or coauthored more than 50 papers recently, and holds more than 70 patents. His research interests include self-organizing networks, wireless networking, and radio transmission technology for wireless communications.
\end{IEEEbiography}

\begin{IEEEbiography}[{\includegraphics[width=1in, height=1.25in,clip, keepaspectratio]{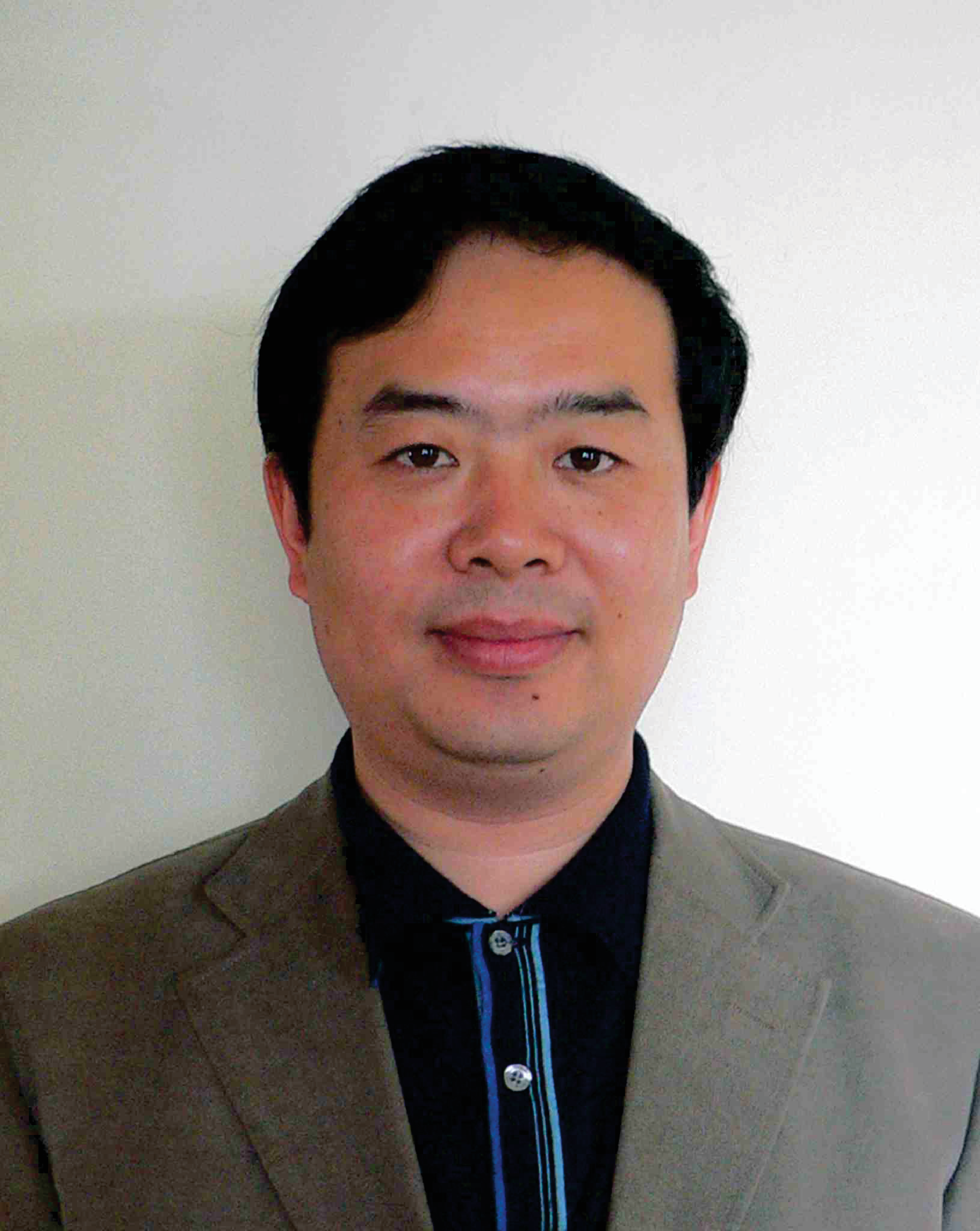}}]{Xiaohu You} (Fellow, IEEE) received his Master and Ph.D. Degrees from Southeast University, Nanjing, China, in Electrical Engineering in 1985 and 1988, respectively. Since 1990, he has been working with National Mobile Communications Research Laboratory at Southeast University, where he is currently professor and director of the Lab. He has contributed over 300 IEEE journal papers and 3 books in the areas of wireless communications. From 1999 to 2002, he was the Principal Expert of China C3G Project. From 2001-2006, he was the Principal Expert of China National 863 Beyond 3G FuTURE Project. From 2013 to 2019, he was the Principal Investigator of China National 863 5G Project. His current research interests include wireless networks, advanced signal processing and its applications.

Dr. You was selected as IEEE Fellow in 2011. He served as the General Chair for IEEE Wireless Communications and Networking Conference (WCNC) 2013, IEEE Vehicular Technology Conference (VTC) 2016 Spring, and IEEE International Conference on Communications (ICC) 2019.
\end{IEEEbiography}

\end{document}